\newcommand{\ls}[1]     
{\dimen0=\fontdimen6\the=#1\dimen0
 \advance\lineskip.5\fontdimen5\the\lineskip-\dimen0
 \lineskiplimit=.9\lineskip
 \baselineskip=\lineskip     \advance\baselineskip\dimen0
 \normallineskip\lineskip
 \normallineskiplimit\lineskiplimit
 \normalbaselineskip\baselineskip
 \ignorespaces
}

\documentclass[journal]{IEEEtran}

\usepackage{graphics}
\usepackage{epsfig}
\usepackage{graphicx}
\usepackage{subfigure}
\usepackage{textcomp}
\usepackage{multirow}
\usepackage{multicol}
\usepackage{cite}

\usepackage{color, soul}
\usepackage{comment}

\usepackage{algorithm}
\usepackage{algpseudocode}

\algtext*{EndWhile}
\algtext*{EndIf}
\algtext*{EndFor}

\ifCLASSINFOpdf
\else
\fi
\begin{document}

\title{
SEOFP-NET: Compression and Acceleration of Deep Neural Networks for Speech Enhancement Using Sign-Exponent-Only Floating-Points
}

\author{

Yu-Chen Lin,
Cheng Yu,
Yi-Te Hsu,
Szu-Wei Fu,
Yu Tsao,~\IEEEmembership{Senior Member,~IEEE,}
Tei-Wei Kuo,~\IEEEmembership{Fellow,~IEEE}

\thanks{Y.-C Lin is with the Department of Computer Science and Information Engineering, National Taiwan University, Taipei 10617, Taiwan, and also with the Research Center for Information Technology Innovation, Academia Sinica, Taipei 11529, Taiwan (e-mail: f04922077@csie.ntu.edu.tw).}

\thanks{Y.-T Hsu is with the Department of Computer Science, Johns Hopkins University, Baltimore, Maryland, 21287 USA (e-mail: b01901112@ntu.edu.tw).}


\thanks{S.-W. Fu, C. Yu, and Y. Tsao are with the Research Center for Information Technology Innovation, Academia Sinica, Taipei 11529, Taiwan (e-mail: jasonfu@citi.sinica.edu.tw).}


\thanks{T.-W Kuo is with the Department of Computer Science and Information Engineering, National Taiwan University, Taipei 10617, Taiwan, and also with the College of Engineering, City University of Hong Kong, Hong Kong (e-mail: ktw@csie.ntu.edu.tw).}

\thanks{Manuscript received xxx xxx, xxx; revised xxx xxx, xxx.}
}

\markboth{Journal of \LaTeX\ Class Files,~Vol.~xxx, No.~xxx, xxx~xxx}%
{Shell \MakeLowercase{\textit{et al.}}: Bare Demo of IEEEtran.cls for IEEE Journals}

\maketitle

\begin{abstract}

Numerous compression and acceleration strategies have achieved outstanding results on classification tasks in various fields, such as computer vision and speech signal processing.
Nevertheless, the same strategies have yielded ungratified performance on regression tasks because the nature between these and classification tasks differs.
In this paper, a novel sign-exponent-only floating-point network (SEOFP-NET) technique is proposed to compress the model size and accelerate the inference time for speech enhancement, a regression task of speech signal processing.
The proposed method compressed the sizes of deep neural network (DNN)-based speech enhancement models by quantizing the fraction bits of single-precision floating-point parameters during training.
Before inference implementation, all parameters in the trained SEOFP-NET model are slightly adjusted to accelerate the inference time by replacing the floating-point multiplier with an integer-adder.
For generalization, the SEOFP-NET technique is introduced to different speech enhancement tasks in speech signal processing with different model architectures under various corpora.
The experimental results indicate that the size of SEOFP-NET models can be significantly compressed by up to 81.249\% without noticeably downgrading their speech enhancement performance, and the inference time can be accelerated to 1.212$\times$ compared with the baseline models.
The results also verify that the proposed SEOFP-NET can cooperate with other efficiency strategies to achieve a synergy effect for model compression.
In addition, the just noticeable difference (JND) was applied to the user study experiment to statistically analyze the effect of speech enhancement on listening.
The results indicate that the listeners cannot facilely differentiate between the enhanced speech signals processed by the baseline model and the proposed SEOFP-NET.
To the best knowledge of the authors, this study is one of the first research works to substantially compress the size of DNN-based algorithms and reduce the inference time of speech enhancement simultaneously while maintaining satisfactory enhancement performance.
The promising results suggest that the application of DNN-based speech enhancement algorithms with the proposed SEOFP-NET technique is more suitable to light-weight embedded devices.

\end{abstract}


%
\IEEEpeerreviewmaketitle

\section{Introduction}
\label{sec:Introduction}

\IEEEPARstart{I}{n} recent years, many applications in different fields have widely used deep neural network (DNN)-based approaches.
These networks can perform well because their deep structures enable DNN-based algorithms to efficiently derive characteristic features while executing various classification and regression tasks.
Many studies have verified that DNN-based algorithms outperform traditional techniques in different computer vision and speech signal processing domains, such as image detection~\cite{imgRec1, ResNet2016}, object detection~\cite{obj1,obj2,obj3}, speech recognition~\cite{Speech2013,Speech2012,Jinyu2013,ASR3,ASR4,ASR5,ASR6}, speaker recognition~\cite{spk1,spk2,spk3}, and speech enhancement~\cite{SE0,SE1,SE2,SE3,SE4,SE5,SE6,NSE1,NSE2,NSE3,NSE4,Wang1}.
However, owing to their deep structure features, most DNN-based algorithms require large memory spaces and incur high computing costs.
As a result, many hardware companies have developed high-level computing units, such as graphics processing units~\cite{GPU1,GPU2,GPU3}, to satisfy the requirements of memory and computation.
In addition to personal computers and mainframes, researchers have aimed to employ DNN-based algorithms to applications in embedded devices used around people.
In this Internet-of-Things era, the number of small embedded devices has exponentially increased.
Such devices cannot be equipped with large storage and high-level computing units.
That is, the applications in embedded devices can only use DNN-based algorithms by accessing DNN models on remote servers through network connections.
However, the latency or disconnection of wireless communication influences the requirements of real-time predictions.
Accordingly, researchers have attempted to locally install DNN-based algorithms in embedded devices.

To implement DNN-based algorithms in embedded devices, the algorithms must be compressed, and computational costs must be reduced.
To resolve this problem, many researchers have successfully developed various compression methods~\cite{BinaryConnect,G2014,INQ,bridgeC,Wang2,Li1}.
The BinaryConnect algorithm~\cite{BinaryConnect}, which uses 1-bit wide weights in the DNN model, yields satisfactory performance in image classification tasks on various image datasets (e.g., MNIST~\cite{MNIST}, CIFAR-10~\cite{CIFAR10}, and SVHN~\cite{SVHN}).
Gong et al.~\cite{G2014} compressed deep convolutional neural networks (CNNs) using vector quantization.
The primeval weights in the models are replaced by the centroid values through the proposed clustering method.
This method only results in a 1\% loss of classification accuracy for state-of-the-art CNNs.
The incremental network quantization (INQ)~\cite{INQ} converts pre-trained full-precision CNN models into a low-precision version.
The weights in the CNN models are all constrained to be either powers of two or zero.
Several experiments on image classification tasks over different well-known CNN architectures (e.g., AlexNet~\cite{alexnet}, VGGNet~\cite{vggnet}, and GoogleNet~\cite{googlenet}) have been conducted.
The experimental results show that the proposed INQ method achieves slightly better performances in the top-1 and top-5 errors using 5-bit quantization.
Most of these compression methodologies have been clearly observed as implementable in classification-based DNN models, such as image recognition~\cite{G2014,INQ,bridgeC} and speech recognition~\cite{qspeech1,qspeech2,qspeech3,qspeech4,qspeech5,qspeech6,ko2017precision}, which classify the input data into a set of output categories.
In contrast, for regression tasks, the output has continuous values.
In brief, the output form of regression tasks considerably differs from that of classification tasks.
Among the existing techniques, Ko et al.~\cite{ko2017precision} proposed a precision scaling method for neural networks to achieve efficient audio processing.
They conducted several experiments on both speech recognition (classification task) and speech enhancement (regression task).
The experimental results showed that the proposed technique exhibited unsatisfactory performance on speech enhancement but acceptable performance on speech recognition.
Sun et al.~\cite{sun2017optimization} developed an optimization method for DNN-based speech enhancement models by utilizing a weight-sharing technique for model compression.
Their experimental results showed that although the size of the DNN model was compressed, the speech enhancement performance was clearly degraded.
Hence, even though the foregoing techniques can efficiently reduce the DNN model size, they also degrade the model performance on speech signal processing regression tasks, such as speech enhancement.
That is, regression tasks compared with classification tasks are more sensitive to value changes in the parameters of DNN-based models.

In the present work, a novel sign-exponent-only floating-point network (SEOFP-NET) is proposed.
It is a neural network whose parameters are represented by a sign-exponent-only floating-point for model compression and inference acceleration of speech enhancement tasks.
Hence, it is an extremely useful application for speech signal processing.
The proposed SEOFP-NET compresses DNN-based speech enhancement models by quantizing the fraction bits of the original single-precision floating-point representation.
After training, all parameters in the trained SEOFP-NET model are slightly adjusted to accelerate the inference time by replacing the floating-point multiplier logic circuit with an integer-adder logic circuit.
For generalization, several experiments were conducted on two important regression tasks in speech enhancement (i.e., speech denoising and speech dereverberation) with two different model architectures (bidirectional long short-term memory (BLSTM)~\cite{x5,chen2015speech} and a fully convolutional network (FCN)~\cite{FCN})) under two common corpora (TIMIT~\cite{TIMIT1993} and TMHINT~\cite{TMHINT}).
To evaluate the enhancement performance, standardized objective evaluation metrics, including the perceptual evaluation of speech quality (PESQ)~\cite{PESQ2001} and short-time objective intelligibility measure (STOI)~\cite{STOI2011}, were employed.
The experimental results illustrate that the size of the SEOFP-NET model can be substantially compressed by up to 81.249\% without considerably downgrading the enhancement performance.
Moreover, the inference time can be accelerated to 1.212$\times$ compared with that of baseline models.
The result also verifies that our SEOFP quantization can cooperate with other efficiency strategies to achieve a synergy effect.
In addition, for the user study experiment, the just noticeable difference (JND)~\cite{JND1,JND2,JND3} was employed to statistically analyze the effect of speech enhancement on listening.
The experimental results indicate that the listeners cannot effortlessly differentiate between the speech signals enhanced by the baseline model and SEOFP-NETs.
To the best knowledge of the authors, this study is one of the first works that considerably compresses the size of DNN-based algorithms and reduces the inference time of speech enhancement tasks simultaneously while maintaining satisfactory enhancement performance.
These promising results suggest that the application of DNN-based speech enhancement algorithms to various lightweight embedded devices using the proposed SEOFP-NET technique is advantageous.

The remainder of this paper is organized as follows.
In Section~\ref{sec:background}, the background knowledge on speech enhancement and floating-point-based parameters of DNN-based algorithms is first presented; the research motivation is explained thereafter.
Section~\ref{sec:design} elaborates on the size compression of DNN-based models and the acceleration of the inference time of trained models using the proposed SEOFP-NET methodology.
Section~\ref{sec:experiments} describes the conduct of experiments in the study with various datasets and different speech enhancement tasks to illustrate the generalization of the proposed SEOFP-NET algorithm.
The disentanglement measurements with various metrics and the relevance of the SEOFP-NET algorithm to speech enhancement are also presented in this section.
Finally, Section~\ref{sec:conclusion} concludes the paper with discussions and description of future work.
\section{Background and Motivation}
\label{sec:background}

In Sections~\ref{sec:2_1} and~\ref{sec:2_2}, two important speech enhancement tasks in regression of speech signal processing, i.e., speech denoising and dereverberation, are introduced, respectively.
In Section~\ref{sec:2_3}, the background knowledge on the single-precision floating-point representation, which is commonly used in DNN-based models, is discussed.
Next, in Section~\ref{sec:2_4}, the electronic circuits of multiplication operation are introduced based on the aforementioned representation.
Finally, in Section~\ref{sec:2_5}, a preliminary experiment is presented, and the motivation of this work is explained.


\subsection{Speech Denoising}
\label{sec:2_1}

The purpose of speech denoising is to generate an improved speech signal and remove noise from an original speech composed of clean speech and environmental noise.
Traditionally, people use time–frequency magnitude spectrograms to clean noisy speech signals; this is denoted as spectrogram mapping-based speech denoising~\cite{x2,x3,x4,x5,xia2014wiener,kolbk2017speech,fu2017complex,chen2015speech}.
This implies that the raw noisy speech waveform is converted to the magnitude spectrogram of noisy speech before denoising is implemented.
The conversion not only generates the magnitude spectrogram but also the phase of noisy speech.
After denoising, the processed magnitude spectrogram is converted back to waveform.
Most spectrogram mapping-based algorithms facilely use the phase of the noisy speech to rebuild the waveform of the denoised speech.
Recently, a number of researchers have proposed the use of waveform mapping-based techniques~\cite{fu2017raw,x6,x7,x8,TSE1,TSE2,TSE3,TSE4} to denoise speech in the waveform domain without waveform-spectrogram conversion.
Accordingly, the proposed SEOFP-NET is applied to both spectrogram and waveform mapping-based speech denoising algorithms to illustrate the generalization of the proposed technique.

In recent years, DNN models have been widely used in speech denoising tasks.
Generally, a DNN-based speech denoising model comprises two phases: offline training and online inference.
In the offline training phase, numerous noisy speech signals consisting of various clean speech signals and types of noise exist in the training corpus.
These noisy speech signals in the training corpus are alternately supplied to the DNN-based speech denoising model, which then generates enhanced speech based on the original noisy speech.
To determine the difference between the two utterances, various criteria, such as the mean square error (MSE)~\cite{x2,x3,x4,x5}, L1 norm~\cite{x7}, and STOI~\cite{x6}, are selected as measurement standards.
After measurements, all parameters in the DNN-based model are updated according to the evaluated difference.
By contrast, in the online inference phase, noisy speech containing mismatching clean utterances and noise is supplied to the trained denoising model.
Ultimately, the denoising system generates an enhanced speech based derived from noisy speech.
Although the two phases are similar among different DNN-based denoising systems, various models, such as deep denoising autoencoders (DDAEs)~\cite{x2,xia2014wiener}, CNNs~\cite{x4,fu2017complex}, FCNs~\cite{fu2017raw, x6, x7, x8}, and BLSTMs~\cite{x5,chen2015speech}, may be applied.
In this work, to illustrate the generalization of the proposed technique, the SEOFP-NET is primarily applied to two model architectures, BLSTM and FCN.


\subsection{Speech Dereverberation}
\label{sec:2_2}

Speech reverberation is defined as the combination of speech signal and its multiple reflections from objects or surfaces within a given space.
Speech reverberation has been confirmed to cause severely degraded speech quality and intelligibility defects.
Hence, reverberation can considerably affect certain speech-related applications, such as automatic speech recognition~\cite{ASR1,ASR2,ASR3,ASR4,ASR5,ASR6} and speaker identification~\cite{spk1,spk2,spk3}.
Reverberation also considerably affects all listeners, normal and impaired.
For decades, researchers have proposed numerous approaches to address the reverberant issue.
Conventional dereverberation techniques include the minimum mean square error~\cite{MMSE}, least square, beamforming~\cite{bf}, and matched filtering~\cite{mf}.

With the rapid developments in the deep learning methods over the past decade, the application of non-linear approaches for dereverberation tasks has been proposed.
DNNs or deep fully connected networks with direct mapping methods have been proposed to improve speech-related system performance through the learning capabilities of deep structured networks.
Some researchers have proposed that DDAEs~\cite{DDAE1,DDAE2} can be used to recover anechoic speech signals from their reverberated counterpart.
Deep recurrent neural networks and long short-term memory networks are also known to be effective for dereverberation tasks because of their capability to analyze time sequences.


\subsection{Single-Precision Floating-Point Representation in DNN Models}
\label{sec:2_3}

\begin{figure}[ht]
\centering
\includegraphics[width=0.9\columnwidth]{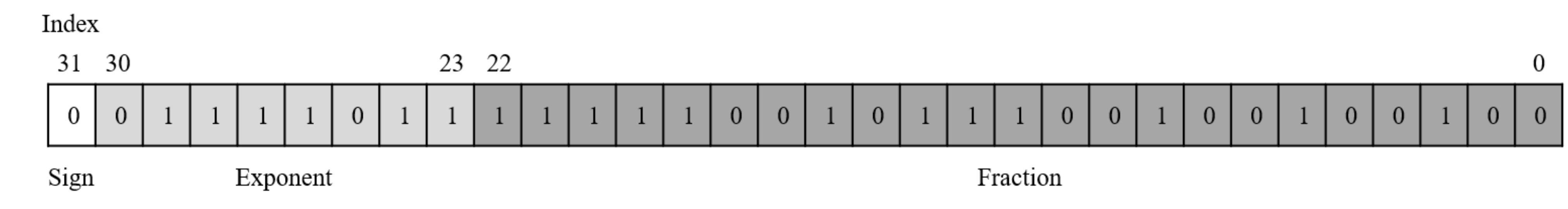}
\caption{An example of the IEEE-754 single-precision floating-point representation. The binary format are divided into three parts: $sign$, $exponent$, and $fraction$. The decimal value of this binary example is 0.123400 corrected to six decimal places.}
\label{fig:binary32}
\end{figure}

The DNN models for either speech denoising or dereverberation contain a considerable number of parameters.
Most of these systems use the IEEE-754 single-precision floating point~\cite{IEEE754} to represent the parameters; Figure~\ref{fig:binary32} shows the binary representation of this floating point.
The binary format has three parts: $sign$, $exponent$, and $fraction$ (also known as significand or mantissa).
The $sign$ part has only one bit, i.e., $bit[31]$, which is regarded as the most important bit in the entire 32-bit binary representation; it indicates the sign of the floating-point value (0 for positive and 1 for negative).
The $exponent$ part has eight bits, i.e., $bit[30-23]$, denoting an unsigned integer, which is the number of times the power of two is raised.
The fraction part has 23 bits, i.e., $bit[22-0]$, representing a real number.
Similar to scientific notation, the decimal value of a single-precision floating point is indirectly calculated using Equation~\ref{eq:FP}.
\begin{equation}
(value)_{10} = (-1)^{sign} \times (fraction)_{10} \times 2^{(exponent)_{10} - bias}.
\label{eq:FP}
\end{equation}
Owing to the unsigned integer feature, a $bias$ is necessary for shifting the value range of the exponent.
In the single-precision floating-point format, the $bias$ for an 8-bit unsigned exponent is 127 ($2^{7}-1$), shifting the value range of the exponent from [0,255] to [-127,128].
In addition, the decimal value of the fraction part can be calculated using the following equation.
\begin{equation}
(fraction)_{10} = 1 + \sum_{i = 0}^{22} bit[i]\times2^{i-23}
\label{eq:fraction-equation}
\end{equation}
For instance, consider the binary representation in Figure~\ref{fig:binary32} in which the $sign$ is $0$, the 8-bit $exponent$ is $01111011$, and the 23-bit $fraction$ is $11111001011100100100100$.
It can be calculated using the following equation.
\begin{equation}
value = (-1)^{0} * (1.9744...) * 2^{123-127} \approx 0.123400
\end{equation}
The equation yields the decimal value 0.123400, which is correct to six decimal places\footnote{ For understanding the conversion from binary to decimal value in detail, please consult https://www.exploringbinary.com/floating-point-converter.}.


\subsection{Arithmetic Electronic Circuits}
\label{sec:2_4}

\begin{figure}[h]
\centering
\includegraphics[width=0.9\columnwidth]{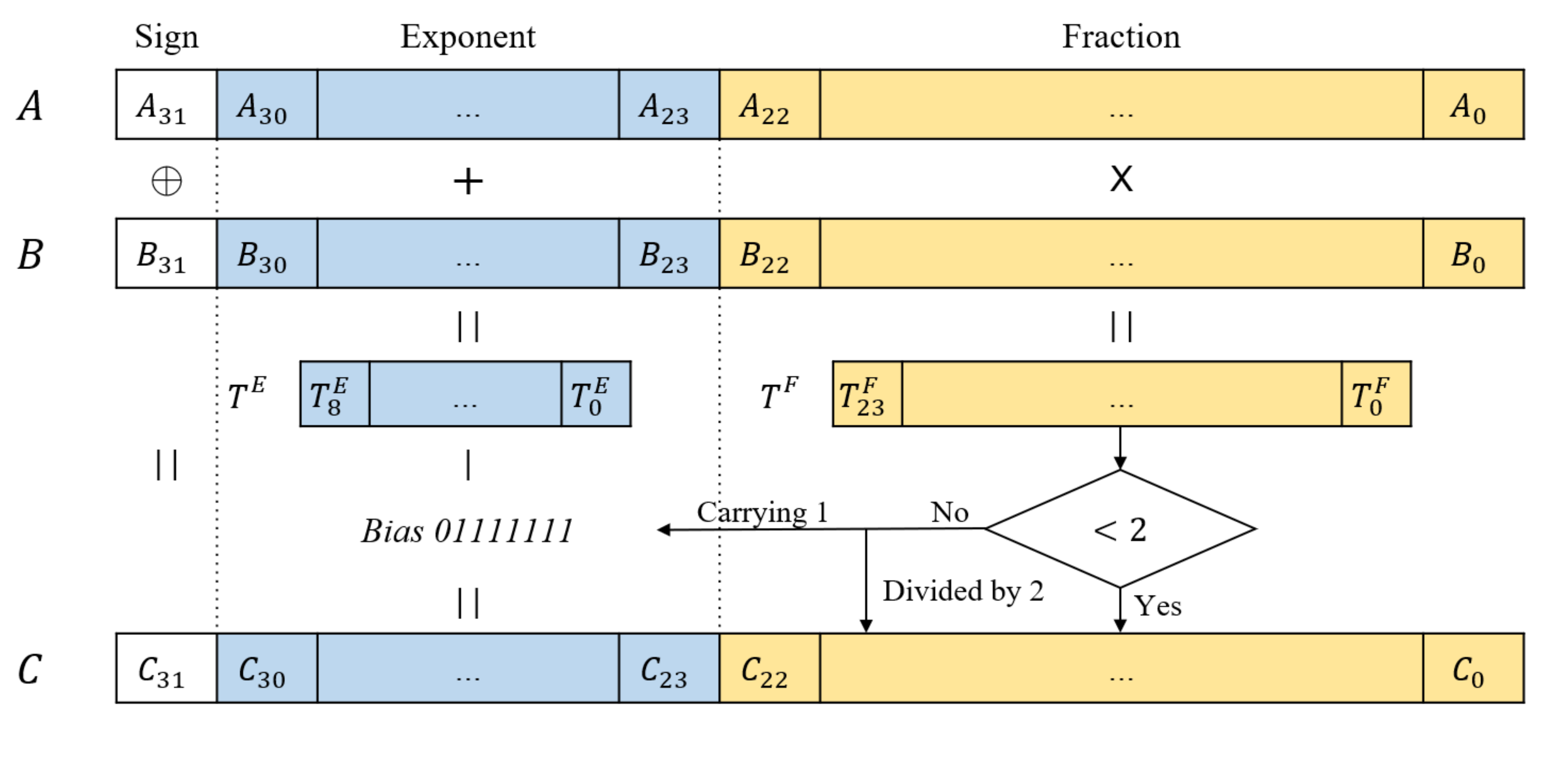}
\caption{Multiplication of two floating-point values $A$ and $B$. The sign bit $C_{31}$ is the result of XOR operation of $A_{31}$ and $B_{31}$. The exponent bits $C_{30-23}$ are the addition result of $A_{30-23}$ and $B_{32-23}$ with consideration of carrying condition from fraction part. The fraction bits $C_{22-0}$ are the multiplication result of unsigned $A_{22-0}$ and $B_{22-0}$.}
\label{fig:mul}
\end{figure}

Figure~\ref{fig:mul} illustrates a single-precision floating-point multiplier circuit that operates the multiplication of two single-precision floating-point values: $A \times B = C$.
For the $sign$ part, two operands, $A[31]$ and $B[31]$, execute an exclusive OR operation to obtain the output sign, $C[31]$, of the result value, $C$.
For the $exponent$ part, two unsigned integers, $A[30-23]$ and $B[30-23]$, first execute an addition operation to obtain a temporary 9-bit output value, $T^{E}[8-0]$
The main reason for the 9-bit width is to cope with $overflow$.
The value range of the addition of two 8-bit unsigned operands whose value range is [0, 255] becomes [0, 510]; consequently, the temporary value, $T^{E}[8-0]$ requires at least 9 bits to receive the output.
The $bias$ of the single-precision floating-point mentioned in Section~\ref{sec:2_3}, i.e., 127, is then subtracted from the temporary output value, $T^{E}[8-0]$.
Considering $32.0 \times 8.0 = 256.0$ as example, the exponent of $256.0$ is $10000111$ which is calculated by the following:
\begin{equation}
10000111 = 10000100 + 10000010 - 01111111
\end{equation}
where $10000100$, $10000010$ and $01111111$ are the binary exponents, $32.0$ and $8.0$, and the single-precision floating-point $bias$, respectively.
For the fraction part, two 23-bit values, $A[22-0]$ and $B[22-0]$, first execute a multiplication operation to obtain a temporary 24-bit output value, $T^{F}[23-0]$.
Similar to the $exponent$ part, the reason for the 24-bit width is to cope with $overflow$.
According to Equation~\ref{eq:fraction-equation}, the original value range of the fraction part is [1, 2); however, after multiplication, the value range becomes [1, 4).
As a result, the temporary value, $T^{F}[23-0]$, requires 24 bits to receive the output.
An if-else decision process determines whether or not $T^{F}[23-0]$ is less than 2.
If it is not less than 2, then $T^{F}[23-0]$ is divided by 2.
The quotient is then carried to the result exponent, $C[30-23]$, and the remainder is considered as the result of the 23-bit fraction, $C[22-0]$.


\subsection{Preliminary Experiment and Motivation}
\label{sec:2_5}

Among the three parts of the single-precision floating-point format, the $sign$ and $exponent$ segments are clearly designed for the value range of the floating-point, and the $fraction$ segment is designed for the value precision.
In the past, precision has been a critical problem for many applications.
Most bits in the single-precision floating-point representation are used for the fraction part (i.e., 23 bits in a 32-bit binary value).
Recently, the single-precision floating-point format has been used in emerging DNN-based algorithms, as mentioned in Section~\ref{sec:2_3}.
However, the necessity of this format to DNN-based speech enhancement algorithms has to be ascertained.
Accordingly, a preliminary experiment is conducted on a BLSTM-based denoising system; the experimental results are summarized in Table~\ref{tab:motivation}.
All parameters of the BLSTM model are directly masked by several 0s at the end.
Two bit lengths for the mask are used: 6 and 12.
The row of the 0-bit mask represents the original denoising model with unmasked parameters.
Based on the list in the table, the downgrade of the PESQ metric scores compared with those of the original model was not distinctly observed.
In addition, these scores increased from a 6-bit mask to a 12-bit mask.
These results motivated the authors to quantize the fraction part of all single-precision floating-point parameters of DNN-based speech enhancement systems.

\begin{table}[h]
	\caption{A preliminary experiment on DNN-based speech denoising under three kinds of bit-width in the fraction of the model parameters. The mask bit-length represents the number of bits that are masked by 0s in the end of the fraction.}
	\centering
	\begin{tabular}{c c c c c }
		\hline
		Mask & \multirow{2}{*}{Binary} & \multirow{2}{*}{Decimal} & \multirow{2}{*}{PESQ}\\
		Bit-length & & & & \\
		\hline
		0 & 0011\ldots1100100100100 & 0.123400002718\ldots & 2.1435\\ 
		6 & 0011\ldots1100100\textbf{000000} & 0.123399734497\ldots & 2.1352\\ 
		12 & 0011\ldots1\textbf{000000000000} & 0.123382568359\ldots & 2.1413\\ 
		\hline 
	\end{tabular}
	\label{tab:motivation}
\end{table}

\begin{figure*}[ht]
	\centering
	\includegraphics[width=2\columnwidth]{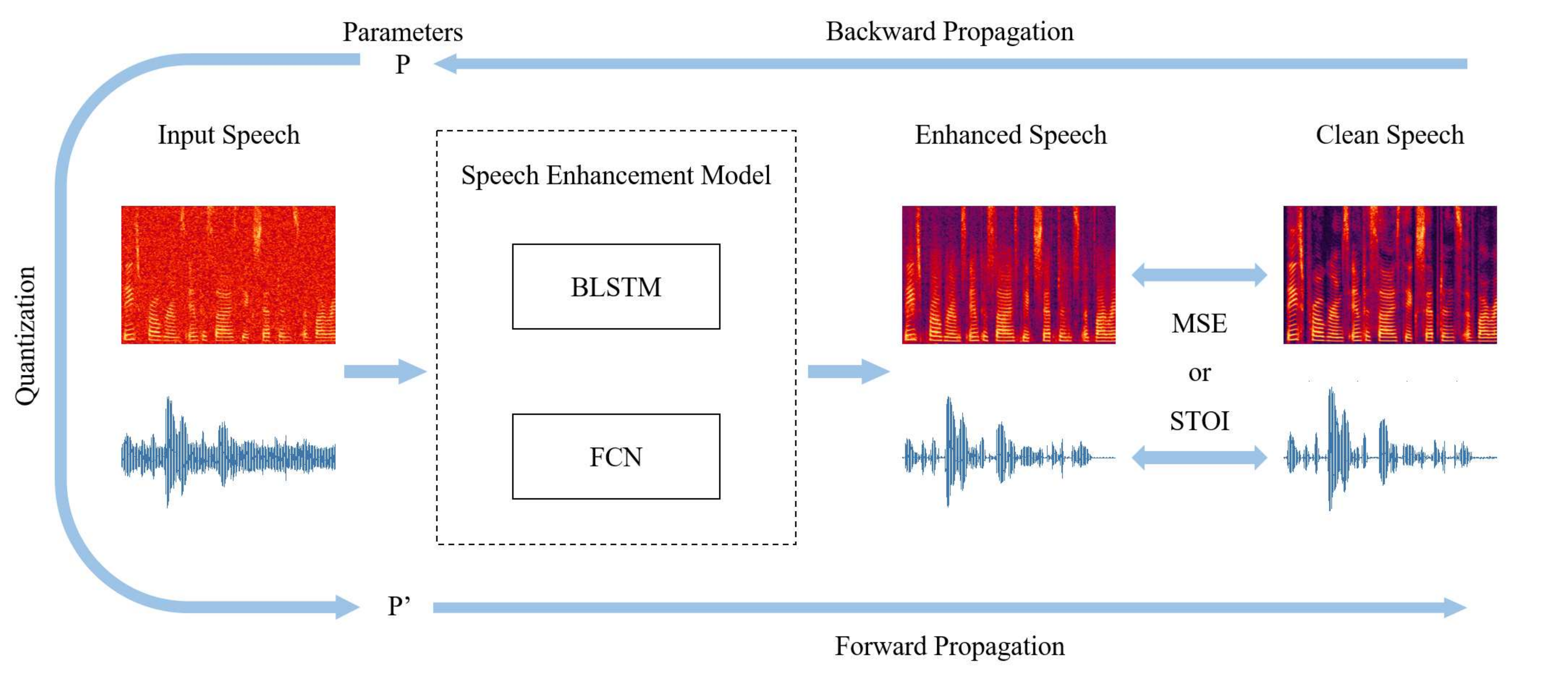}
	\caption{An overview of the DNN-based speech enhancement systems and the training procedure with the proposed SEOFP-NET technique. For generalization, we tried both magnitude spectrogram and raw-waveform as the input data. We also used BLSTM and FCN to illustrate that our SEOFP-NET can be used in different kinds of the model architectures. For evaluating the loss between the enhanced speech signals and the clean speech signals, we used either MSE or STOI as the objective function after the forward propagation.}
	\label{fig:network}
\end{figure*}

Moreover, for executing arithmetic operations, the electronic circuits can be classified into two types: integer and floating point.
In other words, many circuits, such as adders, subtractors, multipliers, and dividers are designed for integer arithmetic, whereas some circuits are designed for floating-point arithmetic.
According to Section~\ref{sec:2_4} and Figure~\ref{fig:mul}, the floating-point multiplier is composed of several sub-circuits of integer arithmetic.
However, many other floating-point circuits also have the same feature, indicating that the circuits for integers are more efficient than the circuits for the floating-point in executing arithmetic operations.
In addition, among the arithmetic circuits for integers, complicated circuits are designed based on simple yet efficient circuits.
Consider the multiplier and divider as examples.
The integer multiplication is completed by several additions, whereas the integer division is completed by several subtractions.
This feature motivated the authors to use simpler and more efficient arithmetic circuits (e.g., integer adder or subtractor) to execute complicated arithmetic operations (e.g., floating-point multiplication or division) for the online inference phase of DNN-based speech enhancement systems.
Using this method, the inference time of DNN models can be accelerated while performing speech enhancement.

However, several design challenges are raised by this scheme.
First, the quantized limitation of the fraction part of the single-precision floating-point format has to be determined, allowing the quantized DNN models to achieve enhancement performance similar to that of the original single-precision model.
Next, before replacing the complicated and inefficient arithmetic circuits with simpler and more efficient alternatives, the three parts of the floating-point parameter have to be adjusted for the arithmetic results to be the same as those of the original arithmetic operations.

\section{SEOFP-NET}
\label{sec:design}

This section presents the proposed SEOFP-NET technique for DNN-based speech enhancement algorithms.
Section~\ref{sec:3_1} introduces the overall training procedure and model architecture of SEOFP-NET.
Section~\ref{sec:3_2} elaborates on the philosophies and algorithm of fraction quantization; it also presents the quantized limitation of the fraction part of the single-precision floating-point format to avoid the severe degradation of speech denoising or dereverberation performance.
Section~\ref{sec:3_3} expounds on the adjustment of the single-precision floating-point parameters of the models to replace the complicated and inefficient floating-point multiplier with a simpler and more efficient integer adder.
Finally, the quantization of the exponent part after training to further compress the model size is discussed in Section~\ref{sec:3_4}.


\subsection{System Overview of SEOFP-NET Quantization}
\label{sec:3_1}

To quantize the fraction part of single-precision floating-point parameters, several bits at the end of the fraction part may be instinctively masked.
Similar to the method employed in the preliminary experiment (Table~\ref{tab:motivation}, Section\ref{sec:2_5}), this may be performed after training a DNN-based speech enhancement model, such as the post-training quantization in Tensorflow Lite~\cite{Tensorflow}.
However, casually modifying the parameters of a well-trained DNN model will affect either the accuracy of a classification task or the performance of a regression task.
The main reason is that this parameter modification does not take the performance change into consideration.
In other words, this intuitive quantization method may considerably degrade the speech enhancement performance.
quantize model parameters while minimizing the influence of task performance, quantization should be implemented during training.
Accordingly, all parameters of the DNN-based speech enhancement model are forced to use a fixed number of bits in the training phase.

Figure~\ref{fig:network} shows an overview of the DNN-based speech enhancement systems and training procedure with the proposed SEOFP-NET technique.
After backward propagation in ($k$)-th iteration, all single-precision floating-point parameters $P$ are quantized to $P'$ by our proposed SEOFP strategy.
The quantized parameters $P'$ are then used as the new model parameters for forward propagation in the succeeding ($k+1$)-th iteration.
During training, the DNN-based speech denoising or dereverberation models learn minimum loss based on quantized parameters.
In addition, we attempted to use both magnitude spectrogram and raw waveform as input data for the speech enhancement system structure.
To illustrate the generalization of different types of model architectures, BLSTM and FCN are used as speech enhancement models.
The MSE or STOI was also used as the objective function for evaluating the loss between enhanced and ground-truth clean speech signals.


\subsection{Fraction Quantization Algorithm}
\label{sec:3_2}

In Figure 1, 71.875\% of the single-precision floating-point memory space (i.e., 23 bits in a 32-bit width representation) is observed to be allocated to the $fraction$ part.
However, such a high-precision 23-bit long fraction part is unnecessary for the parameters of DNN models.
Hence, in the training phase, the DNN-based speech denoising or dereverberation systems are first quantized in the fraction part of all single-precision floating-point parameters.
The quantization algorithm is placed between backward propagation and forward propagation of two adjacent iterations, as mentioned in Section~\ref{sec:3_1}. Besides, the algorithm is applied to all parameters (including weights and biases) in the DNN-models.

\begin{algorithm}[h]
	\caption{Fraction Quantization}
	\begin{algorithmic}[1]
		\Require
		A model $\Lambda$ with $l$ layers, $\{L_i|i=1, 2, \ldots, l\}$.
		A positive integer $x$ for the width of the valid bits in a single-precision floating-point value.
		
		\Ensure A quantized model $\Lambda'$ with all floating-point parameters in bit-width $x$.
		
		\State $kernel \gets [1_{31}1_{30}\ldots1_{31-x+1}0_{31-x}0_{31-x-1}\ldots0_{0}]$
		\For{each layer $L_i$ in the model $\Lambda$}
		\For {each floating-point parameter $P$ in $L_i$}
			\State Convert $P$ into a 32-bit binary variable $B[31:0]$
			\If {$32>x>9$}
				\State $B[32-x] = B[32-x]$ \textbf{\(\vert \vert\)} $B[31-x]$
			\ElsIf {$x=9$}
				\State $B[30:23] = B[30:23] + B[22]$
			\EndIf
			\State $B[31:0] = B[31:0]$ \& $kernel[31:0]$
			\State Convert $B$ back to $P'$
	\EndFor
	\EndFor
	\State \Return $\Lambda'$
	\end{algorithmic}
	\label{algo:fraction}
\end{algorithm}

Algorithm~\ref{algo:fraction} presents the proposed fraction quantization method.
For the input, the algorithm is assigned two input attributes: 1) a DNN model $\Lambda$ with $l$ layers and 2) a positive integer, $x$, to indicate the remaining number of bits after the fraction quantization.
Please note that $\Lambda$ is the model after the backward propagation in any iteration, $k$.
For the output, a model $\Lambda'$, which is quantized with all floating-point parameters in bit width $x$, is used for the forward propagation in the next iteration, $k+1$.
Before the fraction quantization algorithm is applied, a global binary variable, $kernel$ with a 32-bit width is defined.
The head $x$ bits of the kernel are 1s, and the latter $32-x$ bits are 0s.
For each layer, $L_{i}$, of the model, parameter $P$ is fetched; $P$ is first converted from single-precision floating-point data point into a 32-bit binary variable, $B$.
If $x$ is greater than 9 and less than 32, an OR operation is executed with two operands: $B[32-x]$ and $B[31-x]$; the result then updates the value of bit $B[32-x]$.
An OR operation is used to avoid overflow from the $fraction$ segment to the exponent segment.
Here, $B[22:32-x]$ possibly contains all 1s and results in the domino effect of carrying 1.
By contrast, if $x$ is equal to 9 (i.e., only the $sign$ and $exponent$ parts are left in the single-precision floating-point value), the exponent value $B[30:23]$ is then added to the value of $B[22]$.
The use of rounding arithmetic prevents overflow from the $exponent$ segment to the $sign$ segment.
A floating-point value with an $exponent$ consisting of all 1s (11111111) is an infinite value, $\infty$, that does not appear in DNN-based speech enhancement models.
It is impossible that carrying 1 into the exponent part leads to an overflow to the $sign$ bit; hence, the exponent value, $B[30:23]$, can be directly rounded.
The maximum value of $x$ is 32, which means that quantizing the single-precision floating-point parameters is unnecessary.
Finally, the binary variable, $B$, is masked by the binary $kernel$ and then converted back to the floating-point parameter, $P'$.

In short, after the backward propagation in one iteration, $k$, a rounding-like arithmetic is applied to quantize the fraction segment of the single-precision floating-point parameters in the DNN-based speech enhancement model $\Lambda$.
Then, model $\Lambda'$ with all floating-point parameters in bit width $x$ is employed for the forward propagation in the next iteration, $k+1$.
In addition, after applying the fraction quantization strategy to the proposed DNN-based speech enhancement system, the quantized model, whose parameters are only all composed of the sign and exponent parts (i.e., $x = 9$), is found capable of achieving the denoising or dereverberation performance of the original single-precision model.
That is, quantizing all 23 bits in the fraction part of the single-precision floating-point format is the limitation of the fraction quantization strategy.


\subsection{Replacement of Floating-point Multiplier with Integer Adder in Online Inference}
\label{sec:3_3}

After the offline training of a DNN-based speech enhancement system, accelerating the online inference may be attempted.
Because of the simpler electronic circuit design, more efficient integer adders may be employed to function as floating-point multipliers for executing floating-point multiplication operations.
However, a floating-point value and an integer value considerably differ in their binary formats.
Accordingly, a suitable strategy to solve this problem must be developed for the enhanced utterance to remain unchanged after forward propagation.
More specifically, all floating-point parameters in the trained speech denoising or dereverberation model must be adjusted to guarantee that the binary result from the integer addition has the same value as that from the floating-point multiplication.
The following equation illustrates the target of replacement with adjustment on two floating-point operands, $A$ and $B$:
\begin{equation}
(A')_{2} + (B')_{2} = (A)_{2} \times (B)_{2}
\end{equation}
where $+$ is an integer addition operation, and $\times$ is a floating-point multiplication operation; $A'$ and $B'$ are two integer addition operands adjusted according to $A$ and $B$, respectively.
In addition, because there are three parts (i.e., $sign$, $exponent$, and $fraction$) in the binary format of a floating-point parameter, these three segments are adjusted individually, as follows:

\subsubsection{Sign}

Overflow handling is inherent in integer adder circuits; hence, the mechanism is employed to assume the 
XOR operation of sign in the floating-point multiplication.
As mentioned in Section~\ref{sec:2_3}, the sign value is either 0 or 1, representing a positive or negative floating-point value, respectively.
There are only four sign cases in the XOR operation: $\{0,0\}$, $\{0,1\}$, $\{1,0\}$, and $\{0,0\}$; and the XOR results are $0$, $1$, $1$, and $0$ respectively.
The results of the three cases ($\{0,0\}$, $\{0,1\}$, and $\{1,0\}$) are observed to be the same as those operated by the integer addition.
Hence, two sign operands in these cases can be directly added without any label or modification.
For the $\{1,1\}$ case, the addition result is $10$, and the overflow that occurs is labeled by the integer adder.
The integer adder handles this overflow situation by abandoning $1$ and allowing $0$ to remain.
Therefore, in addition to replacing the XOR operation with addition, the overflow label is removed.
In this way, the sign resulting from integer addition is identical to the result that is yielded by floating-point multiplications in all four cases.

\subsubsection{Fraction}

\begin{figure}[ht]
\centering
\includegraphics[width=0.9\columnwidth]{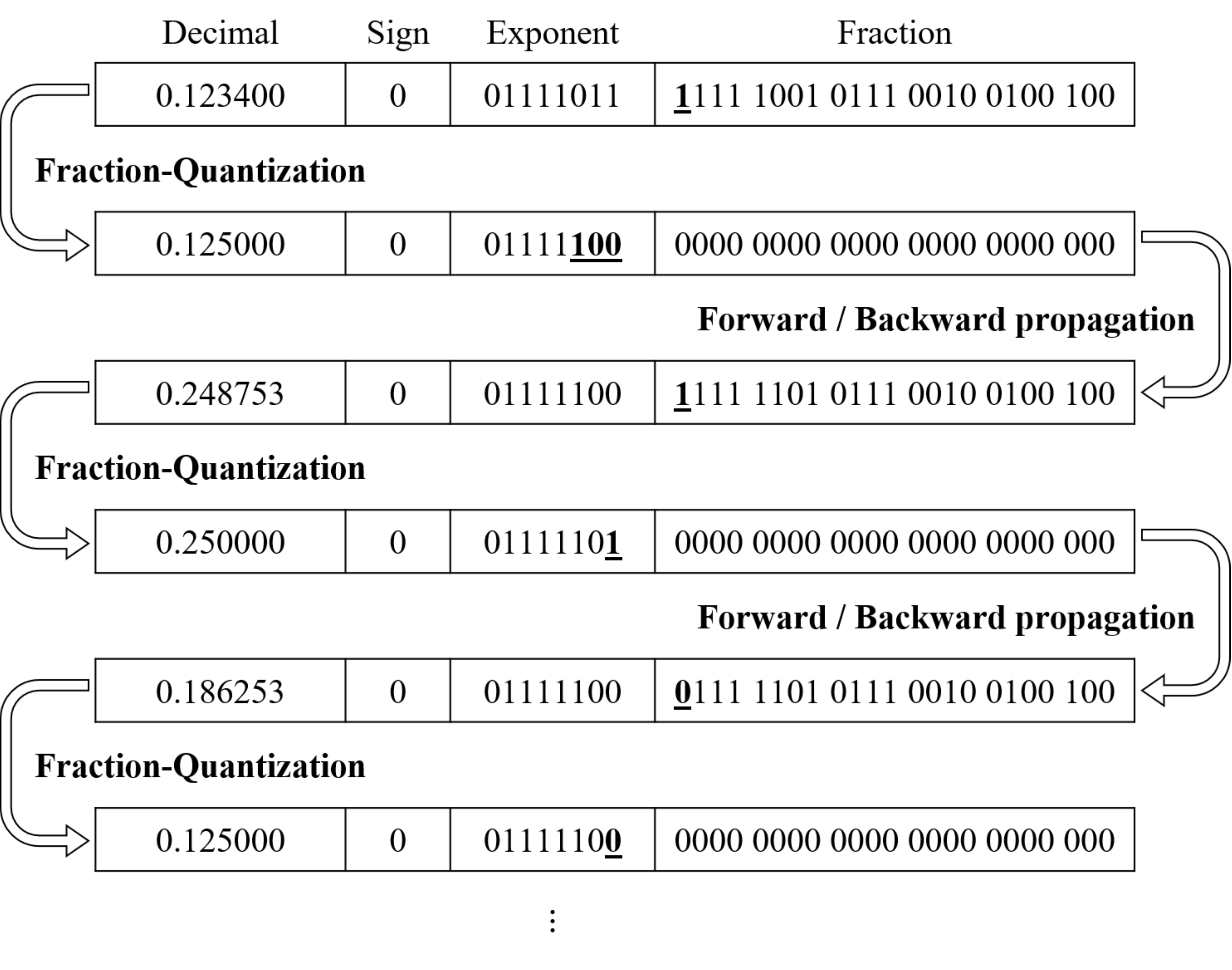}
\caption{An example of the off-line training process for a floating-point parameter $B$. The decimal value of $B$ was 0.123400 in the first iteration. After the rounding-like fraction-quantization, the exponent value of $B$ was carried by 1 and the fraction value became 0. The model then used the updated parameter with value 0.125000 for the next iteration. The algorithm keeps quantizing the model until the end of the training.}
\label{fig:rounding}
\end{figure}

To ensure that the fraction part resulting from the integer addition of $A'$ and $B'$ is the same as that yielded by the floating-point multiplication of $A$ and $B$, the 32 bits in the fraction part of either $A'$ and $B'$ should all be 0s.
In the online inference, because $A$ is represented as an input value of one layer, it is not adjusted; this means that $A'$ has the same value as $A$.
Therefore, only the fraction value of B is adjusted.
The result presented at the end of the previous section indicates that a quantized speech enhancement model may be obtained.
The performance of this model is similar to that of the original single-precision model with parameters composed only of the $sign$ and $exponent$ parts after the offline training.
An example of the offline training process of the floating-point parameter, $B$, is shown in Figure~\ref{fig:rounding}.
With all $B'$ parameters without the fraction values, the fraction resulting from integer addition is identical to the result obtained by floating-point multiplication.

\subsubsection{Exponent}

There are two problems in adjusting the exponent part: the overflow from the exponent segment to the sign segment and the subtraction using the bias, i.e., 127, for the subtractor, as mentioned in~\ref{sec:2_4}.
For the overflow problem, the concept is to avoid the most significant bit (MSB) of the exponent (i.e., the leftmost bit) for both $A$ and $B$ to be 1, allowing the maximum binary value of the exponent segment to become $01111111$.
Consequently, the maximum addition result of the two operands of the exponent is $10000000$, and the sign bit is never affected by the exponent addition.
With this constraint, the bit length of the addition result remains at 8.
To achieve this goal, all input values and model parameters are normalized in the decimal value range $[-1,1]$.
This normalization restricts the binary values of the exponent in the range $[00000000,01111111]$ for the MSB of the exponent to remain 0.
Please note that $00000000$ and $01111111$ are the exponent values of $\pm 0$ and $\pm 1$, respectively.
With this method, the overflow from the exponent segment to the sign segment never occurs.

For the second problem, directly increasing the subtraction after the integer-adder replacement increases the number of instructions.
More specifically, a floating-point multiplication instruction substituted by integer addition and integer subtraction decelerates the online inference.
Accordingly, the subtraction of the bias (127) is divided into two parts: 64 and 63.
The reason for the separation is that both $2^{-64}$ and $2^{-63}$ are decimal values that are extremely smaller than the absolute values of the input and model parameters.
The subtractions of $64$ and $63$ are then distributed to the each of the input values and model parameters.

\begin{figure}[ht]
\centering
\includegraphics[width=0.9\columnwidth]{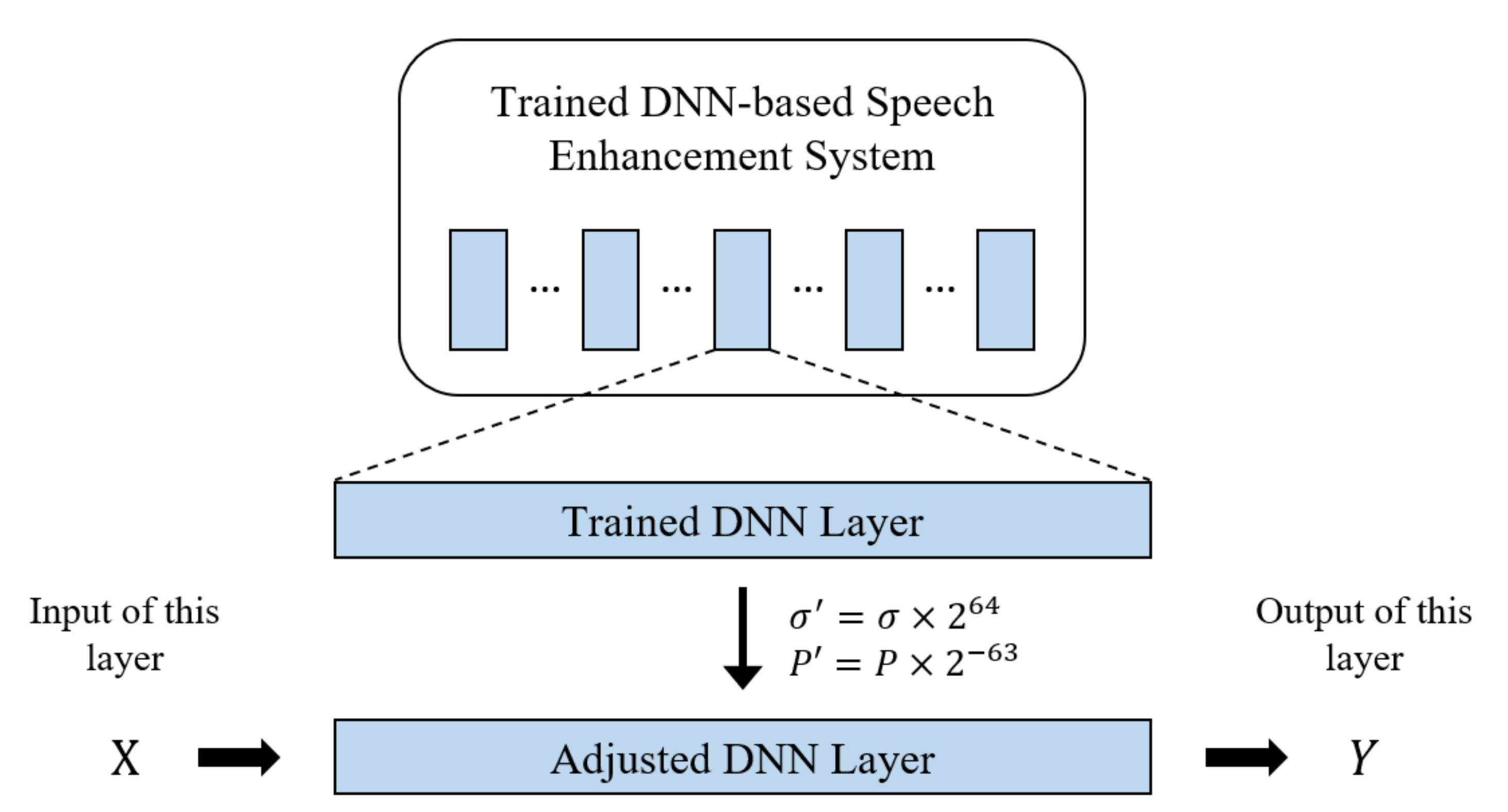}
\caption{Adjustment of a DNN layer in a trained DNN-based speech enhancement system. The model's parameters and the standard deviation of normalization are denoted as $P$ and $\sigma$ respectively.
After the off-line training, the $\sigma$ is multiplied by $2^{64}$ and the parameters $P$ are divided by $2^{63}$.
In the on-line inference, the input $X$ is processed by the adjusted DNN layer with new $P'$ and new $\sigma'$. The output $Y$ is then passed to the next adjusted DNN layer.}
\label{fig:adjustment}
\end{figure}

Figure~\ref{fig:adjustment} illustrates the adjustment of a DNN layer in a trained DNN-based speech enhancement system.
The input and output floating-point values of the layer are denoted as $X$ and $Y$, respectively.
In addition, all model parameters are represented by $Ps$, and $\sigma$ indicates the standard deviation of normalization.
After the offline training, all $Ps$ are divided by $2^{63}$, i.e., the exponent values further subtract 63 ($00111111$); in contrast, $\sigma$ is multiplied by $2^{64}$.
Because $\sigma$ is the denominator, the multiplication of $2^{64}$ to the denominator is equal to the division of the entire value by $2^{64}$.
Therefore, the exponent values further subtract 64 ($01000000$) during the normalization.
The input, $X$, is then processed by the adjusted DNN layer with the new $P'$ and new $\sigma'$; and the output, $Y$, is passed to the next adjusted DNN layer.
With this technique, 127 is further subtracted from the exponent values.

\begin{figure}[ht]
\centering
\includegraphics[width=0.7\columnwidth]{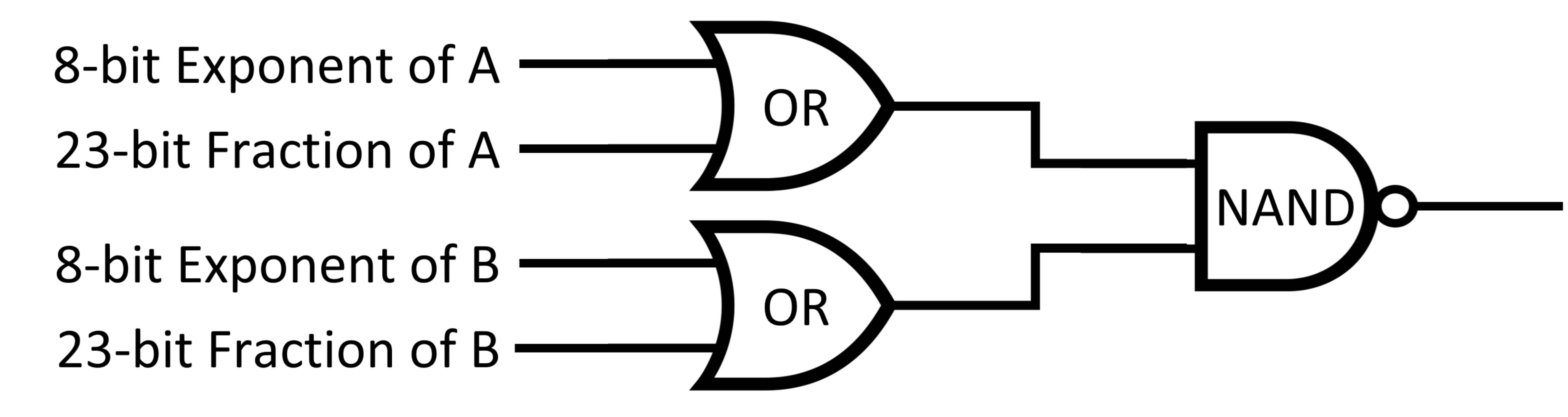}
\caption{An efficient computing logic circuit commonly used for determining whether either of the operands is zero in various computing units. The circuit composes of two OR gates followed by an NAND gate. All 31 bits in the exponent and fraction parts of $A$ and $B$ first execute the OR operations. The results of OR gates then execute a NAND operation. Finally, the result of the NAND gate is 1 indicates the multiplication result is zero since either of the operands is zero.
}
\label{fig:zero}
\end{figure}

\begin{figure}[ht]
\centering
\includegraphics[width=0.9\columnwidth]{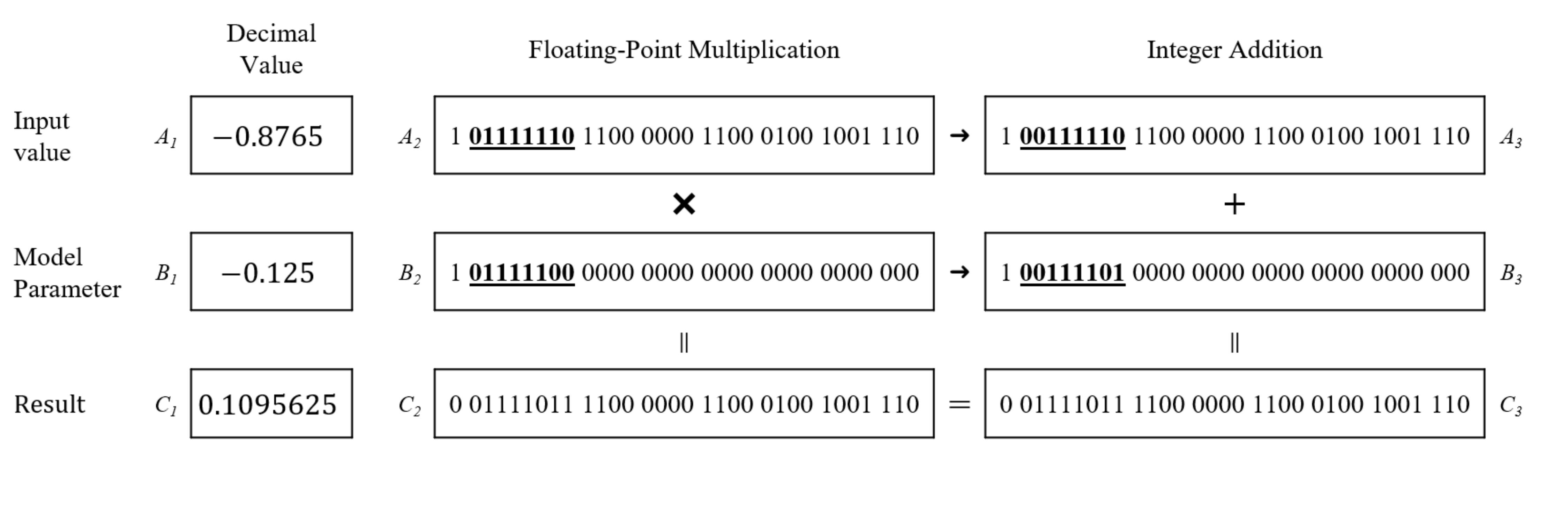}
\caption{An example of the conversion of two floating-point multiplication operands for replacing the floating-point multiplier with a integer adder. The decimal value of two floating-points: \emph{input value} and \emph{model parameter} are $-0.8765$ ($A_{1}$) and $-0.125$ ($B_{1}$); the decimal multiplication result is $0.1095625$ ($C_{1}$).
The binary values of two operands are $A_{2}$ and $B_{2}$.
After the adjustment, the binary values transforms to $A_{3}$ and $B_{3}$.
The multiplication result of $A_{2}$ and $B_{2}$ by a floating-point multiplier is $C_{2}$ which equals to $C_{3}$ the addition result of $A_{3}$ and $B_{3}$ by an integer adder.
}
\label{fig:proof}
\end{figure}

A special case exists in the floating-point multiplication, i.e., the zero-operand multiplication.
If either of the operands is zero, the result of the floating-point multiplication is zero.
For the single-precision floating-point representation, there are two signed zeros, $+0$ and $-0$, whose 31 bits of the exponent and fraction parts are all 0s.
Figure~\ref{fig:zero} shows an efficient computing logic circuit, which is composed of two OR operators followed by an NAND operator; it is commonly used for determining whether either of the operands is zero in various computing units.
If one of the OR results is 0 (which means that either $A$ or $B$ is zero), then the result of the NAND gate is 1, indicating that the multiplication result is 0.
Otherwise, if both OR results are 1s, then the result of the NAND gate is 0, indicating that the multiplication result is not 0.
Accordingly, this efficient logic is applied to the circuit in the zero-operand multiplication case.
Figure~\ref{fig:proof} illustrates an example of the conversion of the two floating-point multiplication operands for replacing the floating-point multiplier with an integer adder.

\subsection{Exponent-Quantization Algorithm}
\label{sec:3_4}

\begin{figure}[ht]
	\centering
	\includegraphics[width=0.9\columnwidth]{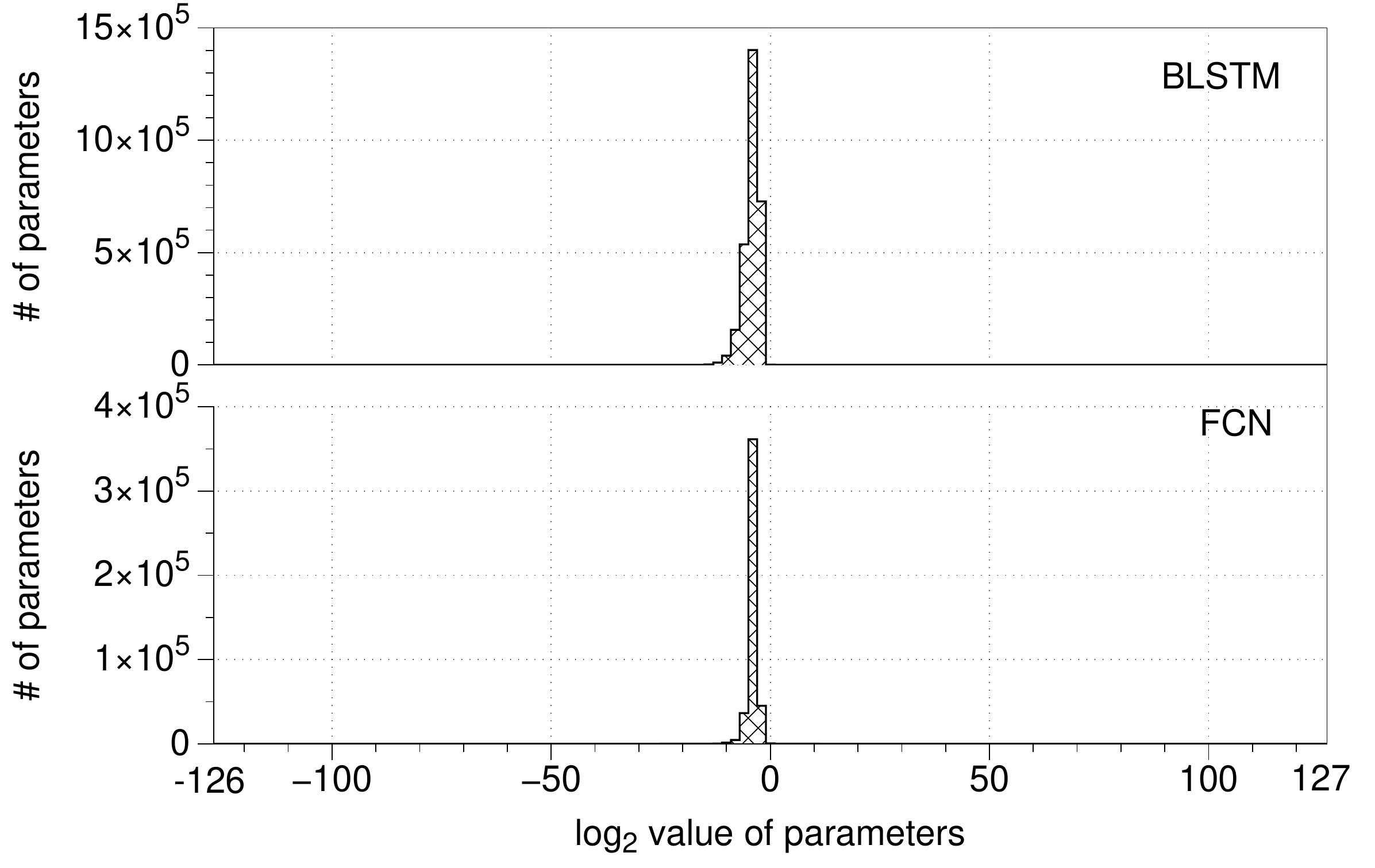}
	\caption{
	Value distribution of all parameters in log$_{2}$ for BLSTM and FCN.
	The x-axis is the log$_{2}$ values and the y-axis is the number of parameters.
	The reason for the range [-126, 127] of the x-axis is that -127 and 128 are used for $0$ and $\infty$ respectively in the 8-bit exponent ranging in [-127, 128].
	}
	\label{fig:exp1distri}
\end{figure}

The maximum number of quantized bits for a single-precision floating-point parameter can easily be verified as 23 based on the fraction quantization algorithm presented in Section~\ref{sec:3_2}.
After the offline training, 9 bits are left in the $sign$ and $exponent$ parts.
Figure~\ref{fig:exp1distri} shows the value distribution of all absolute parameters in the BLSTM and FCN models in log$_{2}$.
Please note that the 8-bit exponent with the bias (i.e., 127) determines an exponent value range in [-127,128], where -127 and 128 are used for 0 and $\infty$.
Thus, the value range in the x-axis in Figure~\ref{fig:exp1distri} is from -126 to 127.
All 9 bits in the $sign$ and $exponent$ parts are known to be designed for a wide value range, i.e., from $\pm2^{-127}$ to $\pm2^{128}$.
However, the normalization process commonly applied to DNN-based models constrains all model parameters within a narrow value range, as shown in Figure~\ref{fig:exp1distri}.
The main reason for applying normalization is to reduce the differences among the parameters.
Based on observation, the parameters are further quantized on these 9 bits.
The exponent part should further be quantized because of the 1 bit in the sign part.
Accordingly, an exponent quantization algorithm based on the value distribution is proposed.

\begin{algorithm}[h]
	\caption{Exponent-quantization}
	\begin{algorithmic}[1]
		\Require A trained model $\Lambda$
		\Ensure  The bit-width $width$. An exponent value $min$ in log$_{2}$. A quantized model $\Lambda'$
		\State Find the $MAX$ and $min$ which are the maximum and minimum decimal exponent values in $\Lambda$ (except for 0).
		\State $width$ = $Ceil$(log$_{2}((MAX-min+1)+1)$)
		\For {each parameter $P$ in $\Lambda$}
			\State $E$ is the exponent value of $P$ in decimal.
			\If{$E \neq 0$}
				\State $E'=E-min+1$
			\EndIf
			\State $P'$ with the exponent value $E'$ substitutes for $P$ in $\Lambda'$
		\EndFor

		\State \Return $width$, $min$, and $\Lambda'$
%
	\end{algorithmic}
	\label{algo:post}
\end{algorithm}

Algorithm~\ref{algo:post} illustrates the proposed exponent quantization.
For the input, a trained speech enhancement model, $\Lambda$, is afforded to the algorithm.
For the output, three output attributes are considered: 1) $width$ indicating the number of bits necessary for the exponent to represent all parameters; 2) $min$ denoting the minimum exponent value in the model; and 3) a quantized model, $\Lambda'$, for exponent quantization.
First, the $MAX$ and $min$, which are the maximum and minimum exponent values in log$_{2}$ in $\Lambda$ (except for zero values) are determined, respectively.
Thereafter, the least bit width is calculated through the ceiling function of log$_{2}((MAX-min+1)+1)$
The reason for determining the latter is that the $00000000$ binary value is used to represent the zero value, as explained in the previous paragraph.
Next, for each parameter in model $\lambda$, the exponent value, $E$, is fetched from parameter $P$.
If $E$ is not equal to zero, then the new exponent value, $E'$, is calculated by $E-min+1$, i.e., the new exponent value, $E'$, is the offset between E and min.
The addition of 1 to the end is necessary because $min$ is represented by 1.
Finally, with the new exponent value, $E'$, $P$ is substituted for $P$ in model $\Lambda'$.
For instance, if the parameter value range is [$\pm0$, $\pm2^{-11}$, \ldots, $\pm2^{2}$], the $MAX$ and $min$ are $2$ and $-11$ respectively.
In this case, only 5 bits (that is 1+$\lceil$ log$_{2}(2-(-11)+1+1) \rceil$) are required for all parameters.
The first 1 bit is for the sign part.
The new exponent values, i.e., $E's$ of $\pm0$, $\pm2^{-11}$ and $\pm2^{1}$ are 0, 1, and 13 respectively.
Please note that the performance is not affected because the bit length of the exponent part has been reduced without changing the values.

\section{Experiments and results}
\label{sec:experiments}

This section presents the experimental setup and results of SEOFP-NET on the speech denoising and dereverberation tasks.
Section~\ref{sec:setup} first describes the experimental setup including the training/testing datasets, model architectures, and performance metrics.
Sections~\ref{sec:removing} to~\ref{sec:JND} describe the conduct of several experiments to compare the SEOFP-NET technique with other strategies.
Section~\ref{sec:removing} presents the comparison of the proposed rounding-like fraction quantization with a direct removing strategy to show that quantizing the fraction bits without any constraint results in the evident degradation of the speech denoising performance.
Sections~\ref{sec:de-noise} and~\ref{sec:de-reverb} explain the application of the SEOFP-NET technique to the denoising and dereverberation tasks, respectively, and the performance evaluation of enhanced speech signals.
The STOI is also integrated into the objective function to improve the performance of the STOI metric, as discussed in Section~\ref{sec:stoi}.
Section~\ref{sec:acceleration} elaborates on the comparison between the model with the proposed integer adder replacement and the original model in the online inference time.
Section~\ref{sec:exponent} presents the comparison of all the model sizes in the original single-precision floating-point models, models with fraction quantization, and models with fraction-exponent quantization.
In addition, a simple cooperation of our quantization and an existing parameter pruning strategy is presented thereafter.
Finally, the JND metric, which is applied to evaluate the results of the user study, is presented in Section~\ref{sec:JND}.


\subsection{Experimental Setup}
\label{sec:setup}

To clearly demonstrate the capability of the proposed SEOFP-NET, extensive experiments were conducted on different model architectures and datasets.
The SEOFP-NET is also applied to the denoising and dereverberation tasks, demonstrating that the proposed technique can achieve satisfactory performance on different speech enhancement tasks of regression in speech signal processing.
To comprehensively understand the differences in performance among the different speech denoising or dereverberation systems, several metrics are employed to evaluate the quality of the enhanced speech signals.
The datasets, model architectures, and evaluation metrics are detailed as follows:

\subsubsection{Datasets}

In the experiments, the TIMIT corpus~\cite{TIMIT1993} is used as the dataset for the denoising task, whereas the TMHINT corpus~\cite{TMHINT} is employed as the dataset for the dereverberation task.
To evaluate the results objectively, a mismatch of noises, SNR, and room impulse responses (RIRs) between the training and testing sets is intentionally designed.
For the training set of the denoising task, all 4620 utterances from the training set of the TIMIT corpus were used.
These utterances were corrupted with 100 different types of noise that are both stationary and non-stationary at eight different signal-to-noise (SNR) levels (from -10 to 25 dB at steps of 5 dB) to generate $4620\times100(types)\times8(SNRs)=3,696,000$ noisy training utterances.
For the testing set of denoising tasks, 100 utterances from the testing set of the TIMIT corpus were used; these were different from the 4620 utterances employed in the training set.
These utterances were corrupted by five different noise types (engine, street, two talkers, baby cry, and white) at four different SNR levels (from -6 to 12 dB at steps of 5 dB) to generate $100\times5(types)\times4(SNRs)=2000$ noisy testing utterances (i.e., 2.2 hours of noisy testing data).

For the dereverberation task, three room conditions were simulated to generate different acoustic characteristics: room 1, room 2, and room 3 with dimensions $4\times4\times4$ m, $6\times6\times4$ m, and $10\times10\times8$ m, respectively.
For the dereverberation task training set, 360 utterances of the TMHINT corpus were employed.
These utterances were convolved with three different RIRs and three considerations of  $T_{60}$, i.e., 0.3, 0.6, and 0.9 (s) to generate $360\times3(T_{60}s)\times3(RIRs)=3240$ reverberant training utterances (i.e., 3.2 hours of reverberation training data).
For the testing set of dereverberation tasks, 120 utterances from the TMHINT corpus are used; these are different from the 360 utterances used in the training set.
These utterances were convolved with a single RIR along with three considerations of $T_{60}$, i.e., 0.4, 0.7, and 1.0 (s) to generate $120\times3(T_{60}s)\times1(RIRs)=360$ reverberation testing utterances (i.e., 0.4 hours of reverberation testing data).

\subsubsection{Model Architectures}

For generalization, two different model architectures are used, BLSTM and FCN, as shown in Figure~\ref{fig:network}.
For the BLSTM-based speech enhancement systems, the spectrograms of the speech signals were employed as the system input.
The speech signals were first parameterized into a sequence of 256-dimensional log-power spectrum features.
Then, mapping was performed frame-by-frame using the BLSTM model.
This model has two BLSTM layers followed by two fully connected layers.
Each BLSTM layer has 257 nodes, and the first fully connected layer has 300 nodes; the second layer is a fully connected output layer.
The BLSTM architecture is similar to that employed in~\cite{x5}.
In contrast, for the FCN-based speech enhancement systems, raw-waveform speech signals were directly utilized as the system input/output without further waveform–spectrum conversion.
The FCN model has 10 convolutional layers.
Each of the first nine layers has 30 size 55 filters; the last layer has only one size 55 filter.
The FCN architecture is similar to that used in~\cite{x6}.

\subsubsection{Evaluation Metrics}

To evaluate the performance of speech denoising and dereverberation, two standardized objective evaluation metrics are used: PESQ~\cite{PESQ2001} and STOI~\cite{STOI2011}.
For PESQ, whose score range is from -0.5 to 4.5, a higher value represents better speech signal quality.
For STOI, whose score range is from 0 to 1, a higher score represents better speech signal intelligibility.
In addition, the JND~\cite{JND1,JND2,JND3} is applied to evaluate the response times of participants in determining the similarity between two enhanced speech signals processed by the original single-precision floating-point model and the proposed SEOFP model.


\subsection{Proposed Rounding-like Strategy versus Direct Removal technique for Fraction Quantization}
\label{sec:removing}

\begin{table*}[ht]
\caption{
Detailed PESQ and STOI scores for BLSTM and FCN using the original single-precision floating-point models and the proposed SEOFP-NETs under specific SNR conditions.
Each score is an average score of three noise types (engine, street, and two talkers).
The score reductions are represented in the percentage from Baseline's scores to SEOFP-NET's scores.
}
\centering 
\label{tab:de-noise}
\setlength{\tabcolsep}{3.6mm}
\renewcommand{\arraystretch}{1.15}
\begin{tabular}{|c| cc | cc | cc | cc | cc |}
	\hline
	\multirow{2}{*}{~}
	& \multicolumn{2}{|c|}{\multirow{2}{*}{Noisy}}
	& \multicolumn{4}{|c|}{BLSTM} & \multicolumn{4}{|c|}{FCN}\\
	\cline{4-11} 
	&\multicolumn{2}{c}{} & \multicolumn{2}{|c|}{ Baseline } & \multicolumn{2}{|c|}{   SEOFP-NET} &
	\multicolumn{2}{|c|}{   Baseline  } &
	\multicolumn{2}{|c|}{   SEOFP-NET  } \\
    \hline
    
		
	SNR(dB) & PESQ & STOI & PESQ & STOI & PESQ & STOI & PESQ & STOI & PESQ & STOI \\ 
	
	\hline
	\multirow{3}{*}{-6} & 1.2232 & 0.5094 & 1.4986 & 0.5676 & 1.4881 & 0.5685 & 1.3814 & 0.5483 & 1.4443 & 0.5384 \\
	& - & - & +0.2754 & +0.0582 & +0.2649 & +0.0591 & +0.1582 & +0.0389 & +0.211 & +0.0290 \\
	& - & - & - & - & -0.70\% & +0.16\% & - & - & +4.55\% & -1.81\% \\
	\hline
	
	\multirow{3}{*}{0} & 1.6218 & 0.6592 & 1.9831 & 0.7280 & 1.9620 & 0.7246 & 1.8427 & 0.7189 & 1.8774 & 0.7001 \\
	& - & - & +0.3613 & +0.0688 & +0.3402 & +0.0654 & +0.209 & +0.0597 & +0.2556 & +0.0409 \\
	& - & - & - & - & -1.06\% & -0.47\% & - & - & +1.88\% & -2.62\% \\
	\hline
	
	\multirow{3}{*}{6} & 2.0161 & 0.7996 & 2.3932 & 0.8315 & 2.3612 & 0.8314 & 2.3036 & 0.8403 & 2.2813 & 0.8144 \\
	& - & - & +0.3771 & +0.0319 & +0.3451 & +0.0318 & +0.2875 & +0.0407 & +0.2652 & +0.0148 \\
	& - & - & - & - & -1.34\% & -0.01\% & - & - &  -0.97\% & -3.08\% \\
	\hline
	
	\multirow{3}{*}{12} & 2.4394 & 0.9005 & 2.6991 & 0.8846 & 2.6383 & 0.8842 & 2.7291 & 0.9112 & 2.7003 & 0.8783 \\
	& - & - & +0.2597 & -0.0159 & +0.1989 & -0.0163 & +0.2897 & +0.0107 & +0.2609 & -0.0222 \\
    & - & - & - & - & -2.25\% & -0.05\% & - & - &  -1.06\% & -3.61\% \\
	\hline
	
	\multirow{3}{*}{Average} & 1.8251 & 0.7172 & 2.1435 & 0.7529 & 2.1124 & 0.7522 & 2.0642 & 0.7547 & 2.0758 & 0.7328 \\ 
	& - & - & +0.3184 & +0.0357 & +0.2873 & +0.0350 & +0.2391 & +0.0375 & +0.2507 & +0.0156 \\
	& - & - & - & - & -1.45\% & -0.09\% & - & - &  +0.56\% & -2.90\% \\
	\hline		
\end{tabular}
\end{table*}

\begin{table}[ht]
	\caption{PESQ scores of enhanced speech signals from BLSTM and FCN using the proposed fraction-quantization algorithm and directly removing within 6 different bit-widths on de-noise task.
	} 
	\centering 
	\setlength{\tabcolsep}{3mm}
	\renewcommand{\arraystretch}{1.15}
	\begin{tabular}{|c|c|c|c|c|} 
		\hline
		 & \multicolumn{2}{|c|}{BLSTM} &  \multicolumn{2}{|c|}{FCN} \\
		\hline
		Bit- & Fraction & Directly & Fraction & Directly \\
		width & Quantization & Removing & Quantization & Removing\\
		\hline
		32 & 2.1435 & 2.1435 & 2.0642 & 2.0642 \\
        \hline
        26 & 2.1364 & 2.1352 & 2.0743 & 2.0636 \\
        \hline
        20 & 2.1252 & 2.1413 & 2.0811 & 2.0743 \\
        \hline
        14 & 2.1354 & 2.1364 & 2.0931 & 2.0857 \\
        \hline
        10 & 2.1541 & 2.1455 & 2.0544 & 2.0345 \\
        \hline
        9 & 2.1124 & 2.0975 & 2.0758 & 1.8595 \\
		\hline
	\end{tabular}
	\label{tab:removing}
\end{table}

An intuitive method to quantize the fraction bits is to maintain the required number of x bits and directly remove the last $32-x$ bits in the fraction segment.
However, the removal of some bits without considering their effect may result in performance degradation.
Accordingly, a rounding-like fraction quantization algorithm is proposed in Section~\ref{sec:3_2}.
The performance of this proposed quantization algorithm is compared with that of the direct removal method in the denoising task.
Table~\ref{tab:removing} summarizes the PESQ scores of the models with six different bit widths (i.e., 32, 26, 20, 14, 10, and 9).
Please note that the 32-bit models are the original single-precision floating-point models, and the 9-bit models indicate that all parameters in those models do not have fraction bits.
Each PESQ score listed in Table~\ref{tab:removing} is the average score in the three noise types and four SNR levels.

The PESQ scores listed in the table are only slightly downgraded when the proposed rounding-like fraction quantization is applied. For example, although the bit width decreased from 32 to 9, the PESQ scores degraded by only 1.451\% (from 2.1435 to 2.1124) and -0.562\% (from 2.0642 to 2.0758) for BLSTM and FCN, respectively. The negative value, -0.576\%, indicates that the performance of the 9-bit quantized FCN model is better than that of the original single-precision floating-point FCN model. However, the PESQ scores are apparently downgraded if several bits are directly removed from the fraction part. For example, although the bit width decreased from 32 to 9, the PESQ scores degraded by only 2.146\% (from 2.1435 to 2.0975) and 9.916\% (from 2.0642 to 1.8595) for BLSTM and FCN, respectively. The results illustrate the considerable potential of the proposed fraction quantization, which utilizes a rounding-like strategy to quantize the fraction bits for reducing the approximation error.


\begin{figure*}[ht]
    \centering
    
%

    \subfigure[clean spectogram]{
        \includegraphics[width=0.2\linewidth]{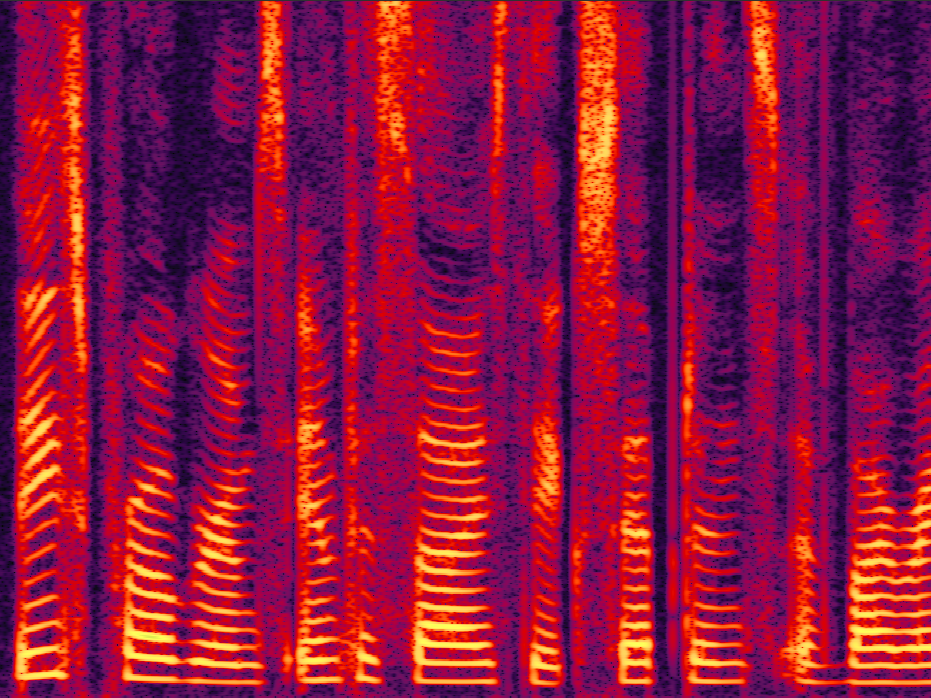}
    }
    \subfigure[noisy spectogram]{
        \includegraphics[width=0.2\linewidth]{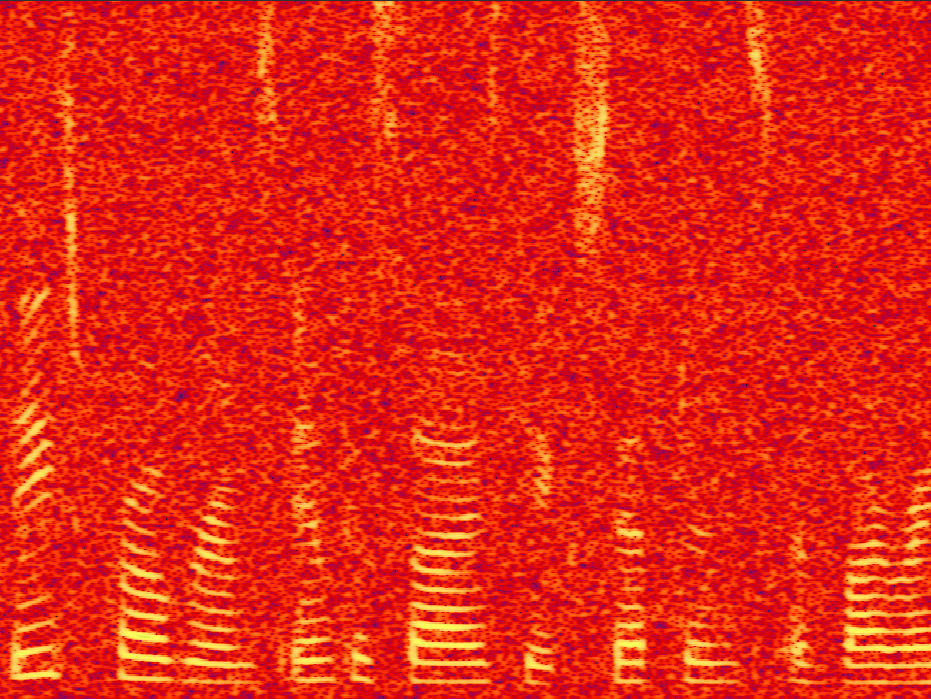}
    }
    \subfigure[Baseline spectogram]{
        \includegraphics[width=0.2\linewidth]{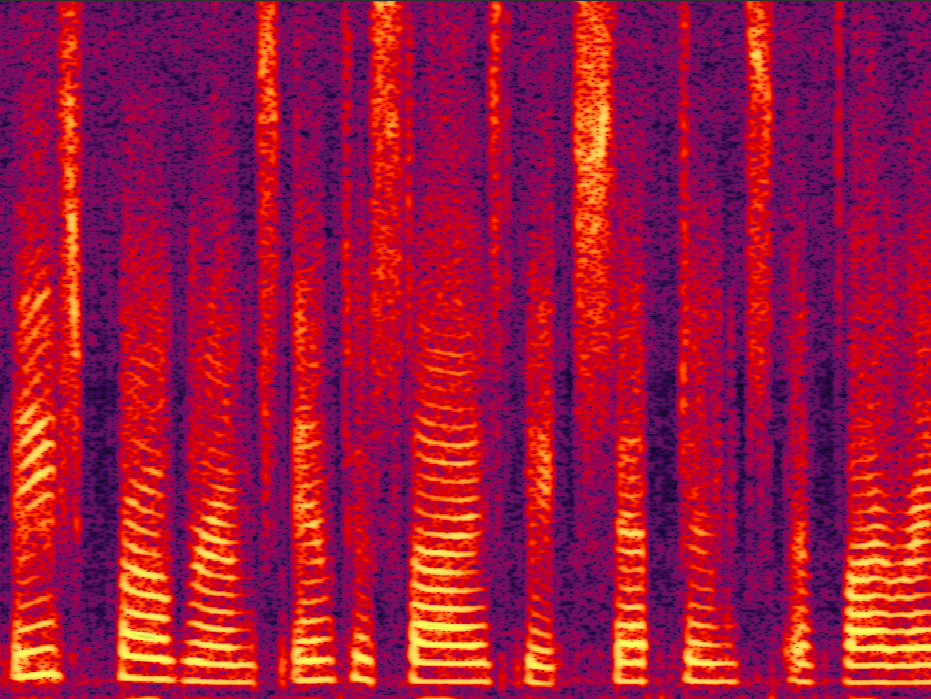}
    }
    \subfigure[SEOFP-NET spectogram]{
        \includegraphics[width=0.2\linewidth]{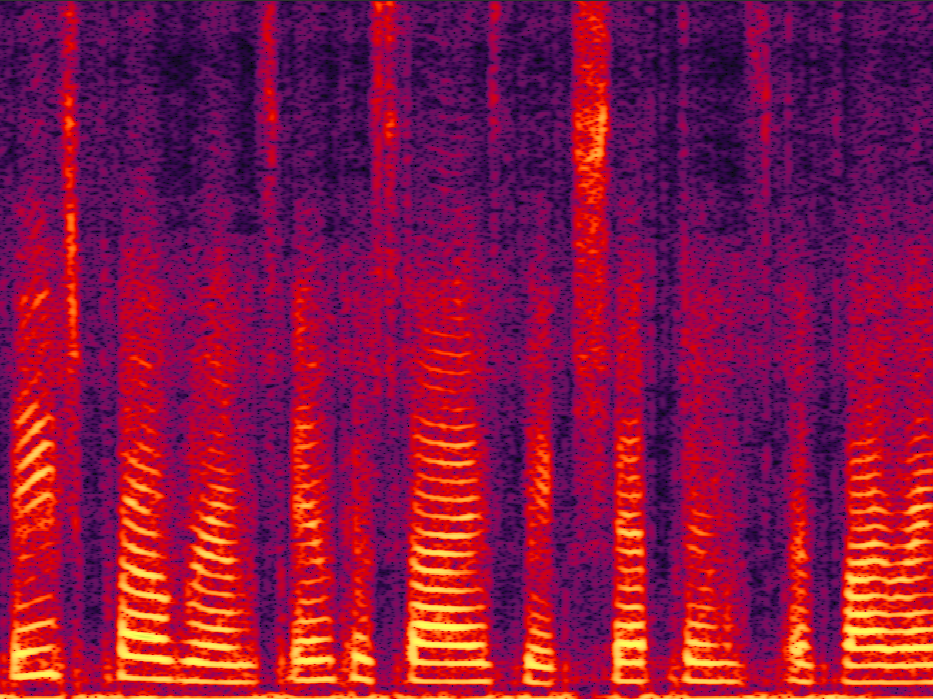}
    }
    
    
    \subfigure[clean raw-waveform]{
        \includegraphics[width=0.2\linewidth]{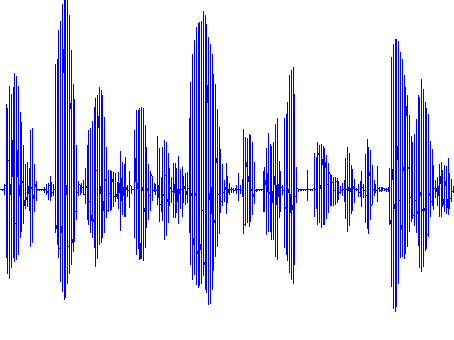}
    }
    \subfigure[noisy raw-waveform]{
        \includegraphics[width=0.2\linewidth]{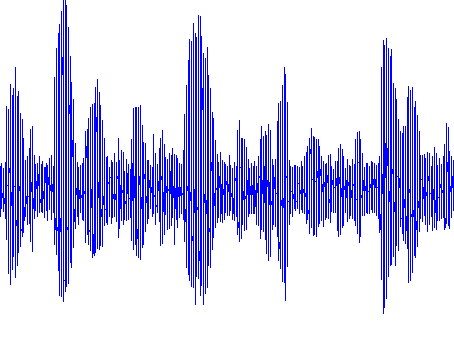}
    }
    \subfigure[Baseline raw-waveform]{
        \includegraphics[width=0.2\linewidth]{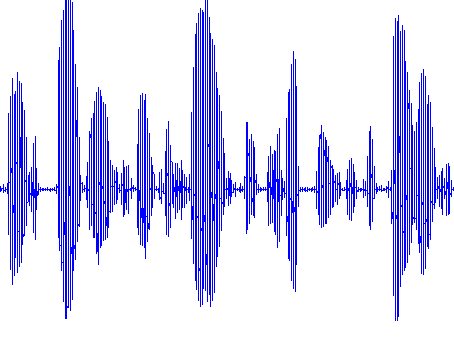}
    }
    \subfigure[SEOFP-NET raw-waveform]{
        \includegraphics[width=0.2\linewidth]{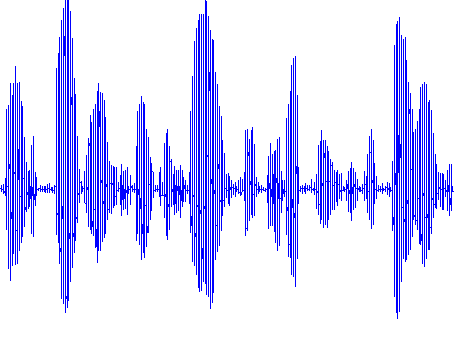}
    }
    
    \caption{
    Spectrograms and waveforms of an example utterance in the de-noise task: (a) and (e) clean speech signals; (b) and (f) noisy speech signals (engine noise); (c) and (g) enhanced speech signals by the original single-precision floating-point FCN; (d) and (h) enhanced speech signals by the proposed SEOFP-NET.
    }
    \label{fig:spec}
\end{figure*}

\subsection{Denoising using Fraction Quantization}
\label{sec:de-noise}

The results listed in Table~\ref{tab:removing} not only show the feasibility of fraction quantization but also suggest that similar denoising performance can still be maintained with only 9 bits (i.e., the sign and exponent bits) left in all model parameters. Table~\ref{tab:de-noise} lists the detailed PESQ and STOI scores for the BLSTM and FCN using the original single-precision floating-point models and the proposed SEOFP-NET models under four specific SNR conditions. Each value is an average score in five noise types (engine, street, two talkers, baby cry, and white). The SEOFP models are quantized by fraction quantization within a 9-bit width. The score reductions are represented as a percentages from the baseline scores to the SEOFP-NET scores. In addition, the PESQ and STOI scores for the unprocessed noisy testing utterances are also listed in the table.


The list in Table II indicates that in applying the SEOFP-NET strategy to the BLSTM-based denoising models, the reduction in the PESQ score is only 1.45\% (from 2.1435 to 2.1124), whereas that for the STOI score is only 0.09\% (from 0.7529 to 0.7522). Similarly, for the FCN-based denoising models, the reduction in the STOI score is only 2.90\% (from 0.7547 to 0.7328); however, the PESQ score improves by 0.56\% (from 2.0642 to 2.0758). A possible reason for this improvement is that the single-precision floating-point parameters may be extremely precise in that the trained parameters overfit the training utterances.


Another observation from Table~\ref{tab:de-noise} is that the FCN-based denoising model has suffered more reductions in the STOI scores after the fraction quantization. A possible reason for this phenomenon is that the number of parameters in the FCN is considerably smaller than the number of parameters in the BLSTM; consequently, each parameter performs a more important denoising function. The same slight error resulting from the fraction quantization of a parameter differently affects the BLSTM and FCN denoising models. Nevertheless, the compression rates are approximately 3.56$\times$ in both models, indicating that the sizes of DNN-based models could be substantially compressed with slight reductions in the denoising performance. Figure~\ref{fig:spec} illustrates the spectrograms and waveforms of a sample utterance in the denoising task. Figure~\ref{fig:spec}(a) to (d), shows the spectrograms of clean speech, noisy (engine noise), enhanced speech processed by the original single-precision floating-point FCN, and enhanced speech processed by the proposed SEOFP-NET, respectively; and Figure~\ref{fig:spec}(e) to (g), shows the speech signals in the waveform format.



\subsection{Dereverberation with Fraction Quantization}
\label{sec:de-reverb}

\begin{table}[ht]
	\caption{
	Detailed PESQ and STOI scores for de-reverberation on the original single-precision floating-point FCN model and SEOFP-NET under three specific reverberation conditions $T_{60}$.
	}
	\centering 
	\label{tab:de-reverb}
	\setlength{\tabcolsep}{2mm}
	\renewcommand{\arraystretch}{1.15}
	\begin{tabular}{| c | cc | cc | cc |}
        \hline
		 & \multicolumn{2}{|c|}{Reverberant} & \multicolumn{2}{|c|}{Baseline} &
		\multicolumn{2}{|c|}{SEOFP-NET} \\
        \hline
		 $T_{60}$ & PESQ & STOI & PESQ & STOI & PESQ & STOI\\ 
		\hline
		
		\multirow{3}{*}{0.4} & 2.1887 & 0.6168 & 2.3217 & 0.7925 & 2.3153 & 0.7550 \\
		& - & - & +0.133 & +0.1757 & +0.1266 & +0.1382 \\
		& - & - & - & - & -0.28\% & -4.73\% \\
		\hline
		
		\multirow{3}{*}{0.7} & 1.8279 & 0.4728 & 1.9086 & 0.7019 & 1.9336 & 0.6495 \\
		& - & - & +0.0807 & +0.291 & +0.1057 & +0.1767 \\
		& - & - & - & - & +1.31\% & -7.47\% \\
		\hline
		
		\multirow{3}{*}{1} & 1.6531 & 0.4010 & 1.6681 & 0.5908 & 1.7494 & 0.5543 \\
		& - & - & +0.015 & +0.1898 & +0.0963 & +0.1533 \\
		& - & - & - & - & +4.87\% & -6.18\% \\
		\hline
		
		\multirow{3}{*}{Ave.} & 1.8899 & 0.4969 & 1.9661 & 0.6951 & 1.9994 & 0.6529 \\
		& - & - & +0.0762 & +0.1982 & +0.1095 & +0.1561 \\
		& - & - & - & - & +1.78\% & -7.99\% \\
		\hline
	\end{tabular}
\end{table}

To illustrate the capability of speech enhancement tasks of regression in speech signal processing, the proposed SEOFP-NET is applied to a speech dereverberation task. Table~\ref{tab:de-reverb} summarizes the details of the PESQ and STOI scores in the dereverberation by the original single-precision floating-point FCN model and SEOFP-NET under the three specific reverberation conditions of $T_{60}$. The score reductions are represented as a percentage from the baseline scores to the SEOFP-NET scores. The PESQ and STOI scores in the unprocessed reverberant testing utterances are listed in the table. The list indicates that in applying the SEOFP-NET strategy to the FCN-based dereverberation model, the STOI score was only reduced by 7.99\% (from 0.6517 to 0.5996). In contrast, the PESQ score improved by approximately 1.78\% (from 1.8728 to 1.9061). These results confirm that the proposed SEOFP-NET may also be applied to different types of speech enhancement tasks of regression in speech signal processing to compress the model sizes of speech dereverberation systems with only marginal degradations in performance. Another observation is that the PESQ scores improved under most reverberation conditions, i.e., $T_{60}$, whereas the STOI scores were reduced under all reverberation conditions in the considered scenarios.
The main reason is that the objective function used in this dereverberation model is the MSE, which has a higher positive correlation with PESQ than with STOI.
The rounding-like approximation, which maintains the quality of the MSE loss value while quantizing the model in the offline training, results in lower performance degradation in PESQ than in STOI.

\subsection{Integration of STOI Metric into Objective Function}
\label{sec:stoi}

\begin{table}[ht]
	\caption{
	Detailed PESQ and STOI scores for de-noise systems using STOI as the objective function on the original single-precision floating-point FCN model and the proposed SEOFP-NET.
	}
	\centering 
	\label{tab:stoi}
	\setlength{\tabcolsep}{1.8mm}{
	\begin{tabular}{| c | cc | cc | cc |}
        \hline
		 & \multicolumn{2}{|c|}{Noisy} & \multicolumn{2}{|c|}{Baseline-S} &
		\multicolumn{2}{|c|}{SEOFP-NET-S} \\
        \hline
		 SNR(dB) & PESQ & STOI & PESQ & STOI & PESQ & STOI\\ 
		\hline
		
		\multirow{3}{*}{-6} & 1.2232 & 0.5094 & 1.3891 & 0.5946 & 1.3803 & 0.5853 \\
		& - & - & +0.1659 & +0.0852 & +0.1571 & +0.0759 \\
        & - & - & - & - & -0.63\% & -1.56\% \\
		\hline
		
		\multirow{3}{*}{0} & 1.6218 & 0.6592 & 1.8280 & 0.7537 & 1.7511 & 0.7272 \\
        & - & - & +0.2062 & +0.0945 & +0.1293 & +0.0680 \\
        & - & - & - & - & -4.21\% & -3.52\% \\
		\hline
		
		\multirow{3}{*}{6} & 2.0161 & 0.7996 & 2.2649 & 0.8675 & 2.1132 & 0.8414 \\
		& - & - & +0.2488 & +0.0679 & +0.0971 & +0.0418 \\
		& - & - & - & - & -6.70\% & -3.01\% \\
		\hline
		
		\multirow{3}{*}{12} & 2.4394 & 0.9005 & 2.6937 & 0.9288 & 2.4893 & 0.9122 \\
		& - & - & +0.2543 & +0.0283 & +0.0499 & +0.0117 \\
		& - & - & - & - & -7.59\% & -1.79\% \\
		\hline
		
		\multirow{3}{*}{Ave.} & 1.8251 & 0.7172 & 2.0439 & 0.7862 & 1.9335 & 0.7665 \\
		& - & - & +0.2188 & +0.0690 & +0.1084 & +0.0494 \\
		& - & - & - & - & -5.40\% & -2.50\% \\
		\hline
	\end{tabular}
	}
\end{table}

The lists in Table~\ref{tab:de-noise} and \ref{tab:de-reverb} indicate that the improvement of enhanced speech in the STOI metric compared with that in PESQ is unclear. Moreover, the reductions (yielded by the original single-precision floating point FCN model and proposed SEOFP-NET) in STOI exceed the reductions in PESQ. The main reason is similar to the explanation provided regarding the previous experiment. The MSE objective function used for training the speech denoising and dereverberation models has a high positive correlation with PESQ; however, the loss function is not sufficient for the STOI metric. Thus, in this experiment, the STOI metric is integrated into the objective function of denoising models. A framework similar to that used in~\cite{x6} was employed.


Table~\ref{tab:stoi} summarizes the PESQ and STOI scores in the denoising systems using STOI as the objective function in the original single-precision floating-point FCN model and the proposed SEOFP-NET. Compared with the FCN scores listed in Table~\ref{tab:de-noise}, the denoising systems with STOI as the objective function are observed as achieving higher improvements in the STOI scores. Consider the improvement under the -6 dB SNR condition as an example (i.e., second row, -6 dB of FCN in Tables~\ref{tab:de-noise} and \ref{tab:stoi}). The Baseline-S and SEOFP-NET-S models improved by 0.0852 and 0.0759 in the STOI metric, respectively, whereas the baseline and SEOFP-NET models only improved by 0.0389 and 0.0290, respectively. In addition, the score reductions (from the baseline model to the SEOFP model) in STOI are smaller than the score reductions in PESQ. For example, in quantizing the parameters, the reduction in the PESQ score was 5.40\% (from 2.0439 to 1.9335); however, the reduction in the STOI score was only 2.50\% (from 0.7862 to 0.7665). These experimental results not only confirm that the STOI optimization is considerably related to achieving speech intelligibility improvement but also suggest that applying the proposed SEOFP-NET with STOI as the objective function decreases the score reduction in the STOI metric.



\subsection{Acceleration by Replacing Floating-Point Multipliers with Integer Adders for Floating-Point Multiplications}
\label{sec:acceleration}

To evaluate the improvement in the inference time, the procedures of the BLSTM and FCN models are simulated using C language on a personal computer with an Intel(R) Core(TM) i7-6700 3.40-GHz CPU. The $clock$ function in the $time.h$ library was also used to evaluate the inference time for speech denoising. In addition, another 12-layer FCN model was created for comparison. Similar to the previous FCN models, the model has 12 convolutional layers with zero padding for the size to be the same as the input. Each of the first 11 layers consists of 30 size 55 filters, and the last layer has only one size 55 filter.


\begin{table}[ht]
	\caption{
	Average inference times and speed up ratios for BLSTM, FCN-10, FCN-12 using the baseline models and the proposed SEOFP-NETs.
	The inference time (per second of testing utterance) are in units of milliseconds.
	}
	\centering 
	\label{tab:acceleration}
	\setlength{\tabcolsep}{2.4mm}
	\renewcommand{\arraystretch}{1.2}
	\begin{tabular}{| c | c | c | c | c | c |}
        \hline
        \multicolumn{2}{|c|}{\multirow{2}{*}{Model}} & \multirow{2}{*}{Metric} & \multirow{2}{*}{Average} & Inference & Speed Up\\
        \multicolumn{2}{|c|}{} & & & Time (ms) & Ratio\\
        \hline
        \multirow{4}{*}{BLSTM} & \multirow{2}{*}{Baseline} & PESQ & 2.1435 & \multirow{2}{*}{78.711} & \multirow{2}{*}{-} \\
        \cline{3-4}
        & & STOI & 0.7529 &  & \\
        \cline{2-6}
        & SEOFP & PESQ & 2.1124 & \multirow{2}{*}{66.020} & \multirow{2}{*}{1.192$\times$} \\
        \cline{3-4}
        & NET & STOI & 0.7522 &  & \\
        \hline
        
        \multirow{4}{*}{FCN$_{10}$} & \multirow{2}{*}{Baseline} & PESQ & 2.0642 & \multirow{2}{*}{110.691} & \multirow{2}{*}{-} \\
        \cline{3-4}
        & & STOI & 0.7547 &  & \\
        \cline{2-6}
        & SEOFP & PESQ & 2.0758 & \multirow{2}{*}{91.308} & \multirow{2}{*}{1.212$\times$} \\
        \cline{3-4}
        & NET & STOI & 0.7328 &  & \\
        \hline

        \multirow{4}{*}{FCN$_{12}$} & \multirow{2}{*}{Baseline} & PESQ & 2.0583 & \multirow{2}{*}{134.035} & \multirow{2}{*}{-} \\
        \cline{3-4}
        & & STOI & 0.7543 &  & \\
        \cline{2-6}
        & SEOFP & PESQ & 2.1074 & \multirow{2}{*}{110.709} & \multirow{2}{*}{1.211$\times$} \\
        \cline{3-4}
        & NET & STOI & 0.7326 & & \\
        \hline
		
	\end{tabular}
\end{table}

Table~\ref{tab:acceleration} summarizes the average online inference times and speedup ratios for BLSTM, FCN$_{10}$, and FCN$_{12}$ using the baseline models and the proposed SEOFP-NETs. The inference times for noisy testing data with an average of 1.3 hours (i.e., 1200 testing utterances) are given in units of millisecond. The list in the table indicates that the inference times are significantly reduced by SEOFP-NET. The acceleration rates of inference time are 1.192$\times$ (from 78.711 to 66.020), 1.212$\times$ (from 110.691 to 91.308), and 1.211$\times$ (from 134.035 to 110.709) for the BLSTM, FCN$_{10}$, and FCN$_{12}$, respectively. That is, the inference time of DNN-based speech denoising systems can simply be accelerated by the proposed SEOFP-NET strategy without expensive and complicated hardware accelerators.


The list in Table~\ref{tab:acceleration} further indicates that although FCN$_{12}$ SEOFP-NET has two more convolutional layers than the FCN$_{10}$ baseline model, their inference times are virtually the same. More specifically, the average inference time of FCN$_{12}$ SEOFP-NET is approximately 110.691 ms (per second of the testing utterance); this approximates the 110.709-ms average inference time of the FCN$_{10}$ baseline model. However, for the performance metrics, FCN$_{12}$ SEOFP-NET outperforms the FCN$_{10}$ baseline model in PESQ and has a STOI score similar to that of the latter. The result suggests that instead of training the DNN-based speech denoising system with fewer layers, training the SEOFP-NET with more layers can improve performance without increasing the online inference time.



\begin{table*}[ht]
	\caption{
	Number of parameters and the model sizes of the BLSTM and FCN using the baseline single-precision models, SEOFP-NETs with only fraction-quantization, and SEOFP-NETs with both fraction- and exponent-quantization algorithms.
	The model sizes are in units of kilobytes (KB).
	The compression ratios are also listed in the table.
	} 
	\centering 
	\setlength{\tabcolsep}{3.5mm}
    \renewcommand{\arraystretch}{1.5}
	\begin{tabular}{ l c c c c} 
		\hline
		& BLSTM & Compression Ratio & FCN & Compression Ratio \\
		\hline
		\hline
		Number of parameters & 2,877,929 & - & 450,301 & - \\
		Size of the baseline models with single-precision floating-point (KB) & 11,242 & - & 1,759 & - \\
		Size of the SEOFP-NETs with Fraction-Quantization (KB) & 3,162 & -71.873\% & 495 & -71.859\%\\
		Size of the SEOFP-NETs with Fraction-Exponent-Quantization (KB) & 2,108 & -81.249\% & 385 & -78.113\% \\
		\hline
		Number of parameters of SEOFP FCN model with parameter pruning & \multirow{2}{*}{-} & \multirow{2}{*}{-} & 337,725 & \multirow{2}{*}{-83.570\%}  \\
		Size of SEOFP FCN model with parameter pruning (KB) &  &  & 289 & \\
		\hline
	\end{tabular}
	\label{tab:exponent}
\end{table*}

\subsection{Further Model Compression by Exponent Quantization Algorithm and Existing Parameter Pruning Strategy}
\label{sec:exponent}

Table~\ref{tab:exponent} lists the number of parameters and model sizes of the BLSTM and FCN using the original single-precision models, SEOFP-NETs with only the fraction quantization, SEOFP-NETs with both fraction and exponent quantization algorithms, and the combination of SEOFP and parameter pruning.
The sizes of all models are in kilobytes (kB).
The compression ratios, which represent the size reduction percentages from the baseline models to the SEOFP-NETs, are also listed in the table and calculated by the following equation:
\begin{equation}
Compression\ ratio = \frac{S_{SEOFP}-S_{B}}{S_{B}} \times 100\%
\end{equation}
where $S_{SEOFP}$ and $S_{B}$ represent the sizes of baseline models and SEOFP-NETs, respectively.

From the table, the sizes of models with the proposed fraction quantization algorithm compared with those of the baseline models are evidently compressed. The sizes of the SEOFP-NETs with fraction quantization compared with those of the baseline BLSTM and FCN models were reduced to 71.873\% (from 11,242 to 3,162) and 71.859\% (from 1,759 to 495), respectively. These compression ratios may be attributed to the redundant fraction bits in the single-precision floating-point parameters.
To further compress the model sizes, the proposed exponent quantization algorithm was applied to the trained BLSTM-based and FCN-based speech denoising systems, as mentioned in Section~\ref{sec:3_4}.
First, the maximum $MAX$ and minimum $min$ exponent values in log$_{2}$ of each model were determined.
After the exponent quantization, the value set \{$MAX$, $min$, $width$\} of each model were obtained, i.e., \{0, -23, 5\} for the BLSTM-based denoising model, respectively, and \{10, -26, 6\} for the FCN-based denoising model, respectively.
The quantized models were also generated by the exponent quantization algorithm.
As indicated in the table, the sizes of the SEOFP-NETs with the fraction and exponent quantization algorithms compared with those of the SEOFP-NETs with only the fraction quantization were further compressed.

Compared with the sizes of the baseline models, those of the SEOFP-NETs were reduced by approximately 80.249\% (from 11,242 to 2,108) and 78.113\% (from 1,759 to 385) for the BLSTM and FCN, respectively.
In other words, the model sizes of SEOFP-NETs are only one-fifth of the sizes of the baseline models.
Further, the compression ratios may be attributed to the narrow value distribution of all parameter exponents.
In addition, to verify that the proposed SEOFP strategy can cooperate with other efficiency strategies to achieve a synergy effect, we combine our quantization strategy with an existing parameter pruning strategy proposed by~\cite{PP1,PP2}. From the table, the number of parameters in the FCN model reduced from 445,301 to 337,725. The model remaining model size is only 16.43\% (from 1,759 to 289 KB). The result is encouraging for combination of different efficiency methodologies finding the compression limitation of speech enhancement without performance degradation.


\subsection{Just Noticeable Difference}
\label{sec:JND}

\begin{figure}[ht]
\centering
\includegraphics[width=0.9\columnwidth]{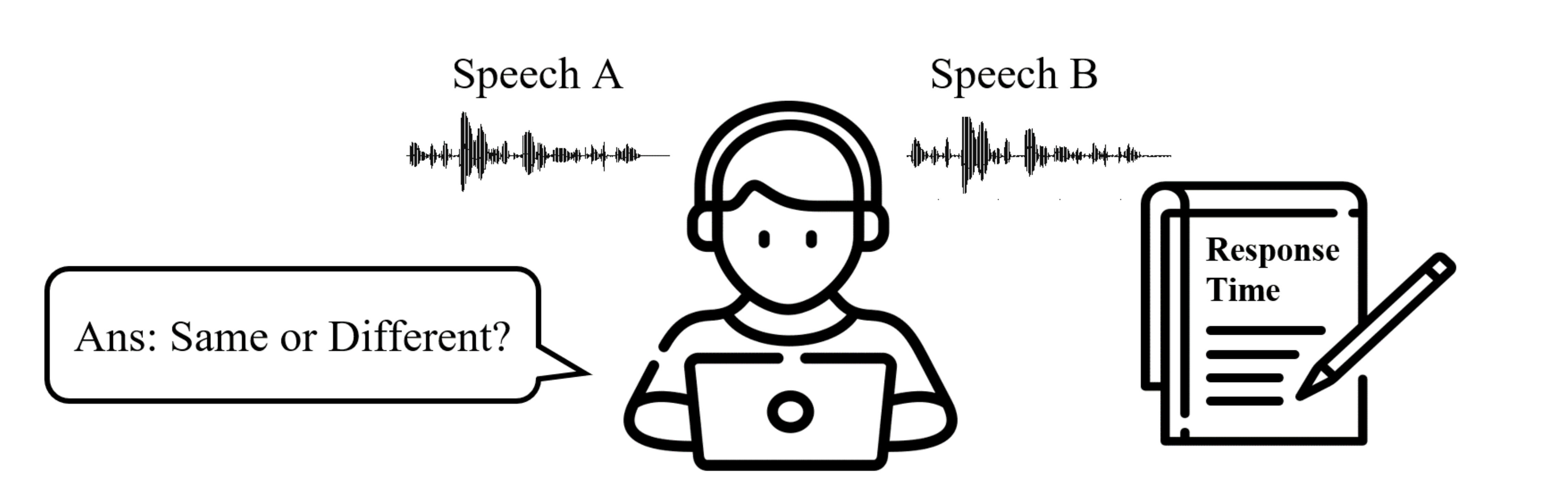}
\caption{
The environment of the JND user study experiment.
For each pair of utterances in the 100-pair A/B Test, the participant listened to two enhanced speech signals and then answer in SAME or DIFF.
The response time was recorded to determine the availability of the response.
}
\label{fig:JND}
\end{figure}

Finally, the JND is employed as the metric for the user study. The JND is the minimum amount of change that can produce a noticeable variation in sensory experience (e.g., sight, hearing, and tactile sense)~\cite{JND1,JND2,JND3}. In the human hearing system, the JND is used to measure the difference among similar speech signals (acoustics or sounds). Accordingly, the JND is an appropriate metric for evaluating the difference between the enhanced speech signals yielded by the baseline model and that of the proposed SEOFP-NET. The environment of this JND user study experiment is shown in Figure~\ref{fig:JND}. In this experiment, six denoising SEOFP-NETs with six different bit widths (i.e., 32, 26, 20, 14, 10, and 9) for the parameters presented in Section~\ref{sec:removing} are employed. For the A/B test for 20 participants, 100 pairs of speech signals were prepared; each participant was requested to listen to the same 100 pairs of speech signals. For each pair, one speech was the utterance enhanced by the baseline model, and the other was enhanced by one of the six SEOFP-NETs. Note that if the latter speech is enhanced by the SEOFP-NET with a 32-bit width, these two speech signals will be the same because the SEOFP-NET with a 32-bit width is exactly the same as the baseline model. Moreover, both speech signals are under the same SNR condition. After listening to these pairs of speech signals, each participant had to determine whether the speech signals were the same or different by giving the answers $SAME$ or $DIFF$, respectively. The response time for the determining the similarity or difference between the two enhanced speech signals was recorded.

\begin{figure*}[ht]
	\centering
	
	\subfigure[amount of DIFF responses]{
        \includegraphics[height=3.8cm]{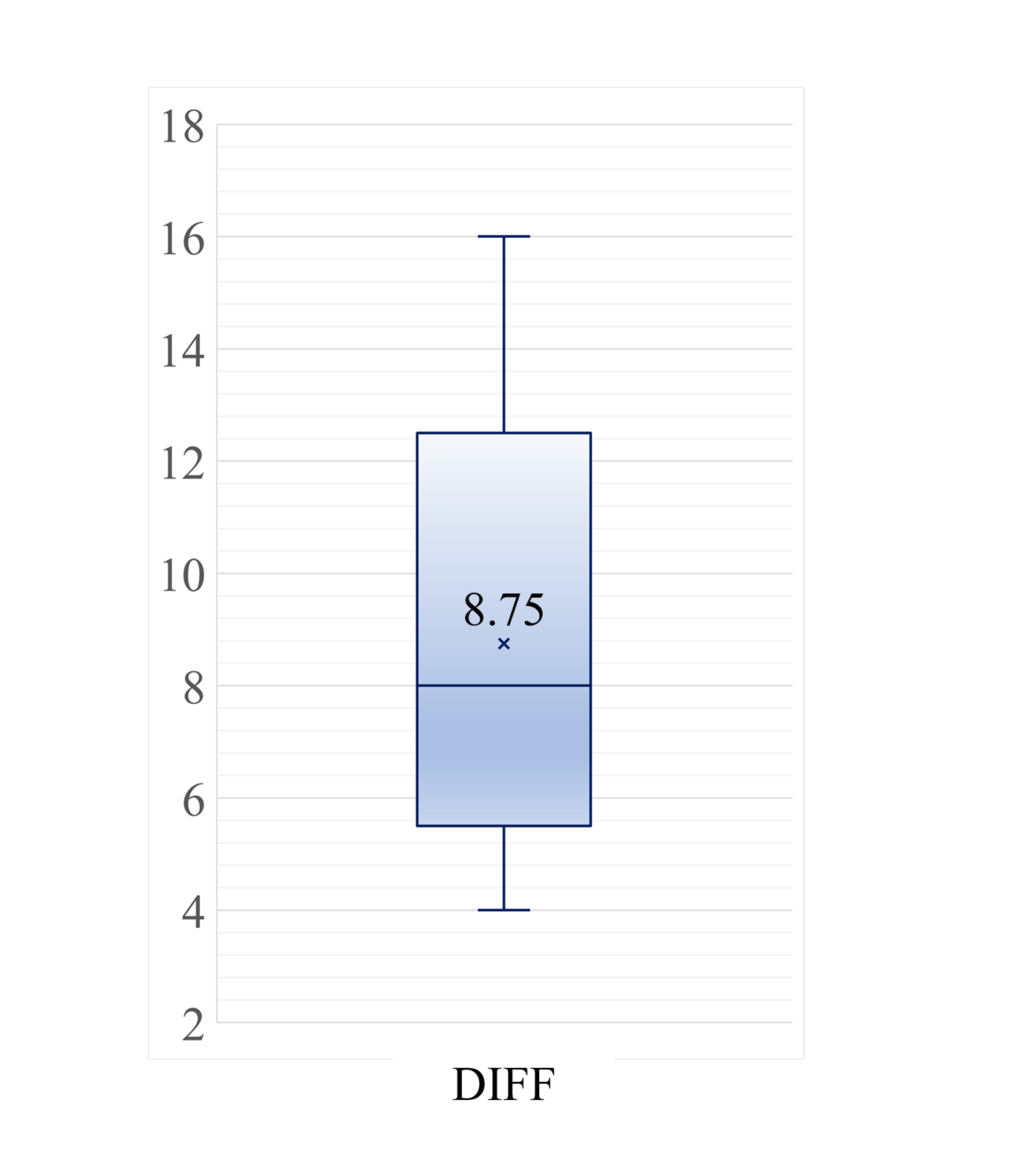}
        \label{fig:JND_a}
    }
    \subfigure[response times]{
        \includegraphics[height=3.8cm]{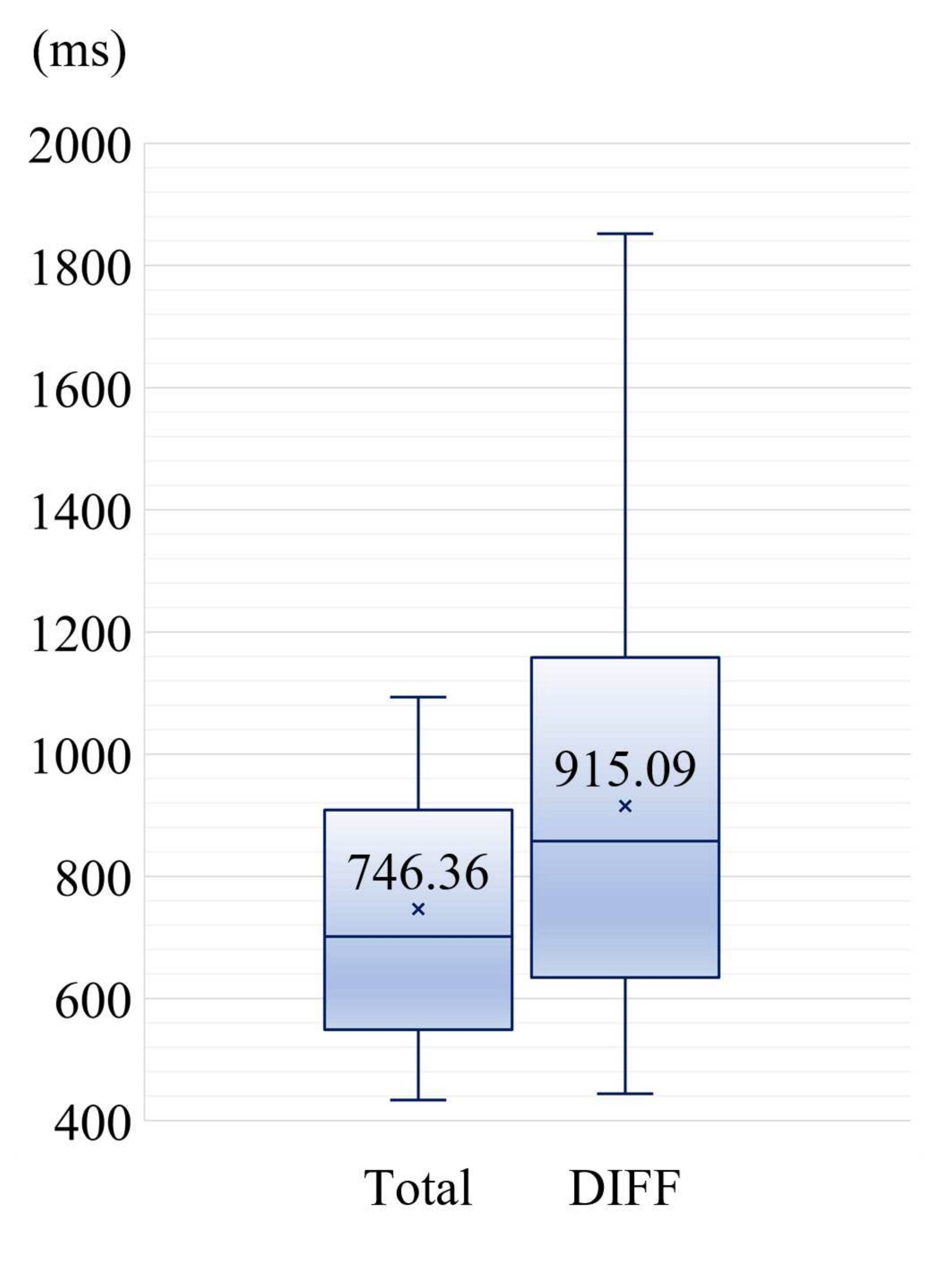}
        \label{fig:JND_b}
    }
    \subfigure[amounts of DIFF responses under different conditions]{
        \includegraphics[height=3.8cm]{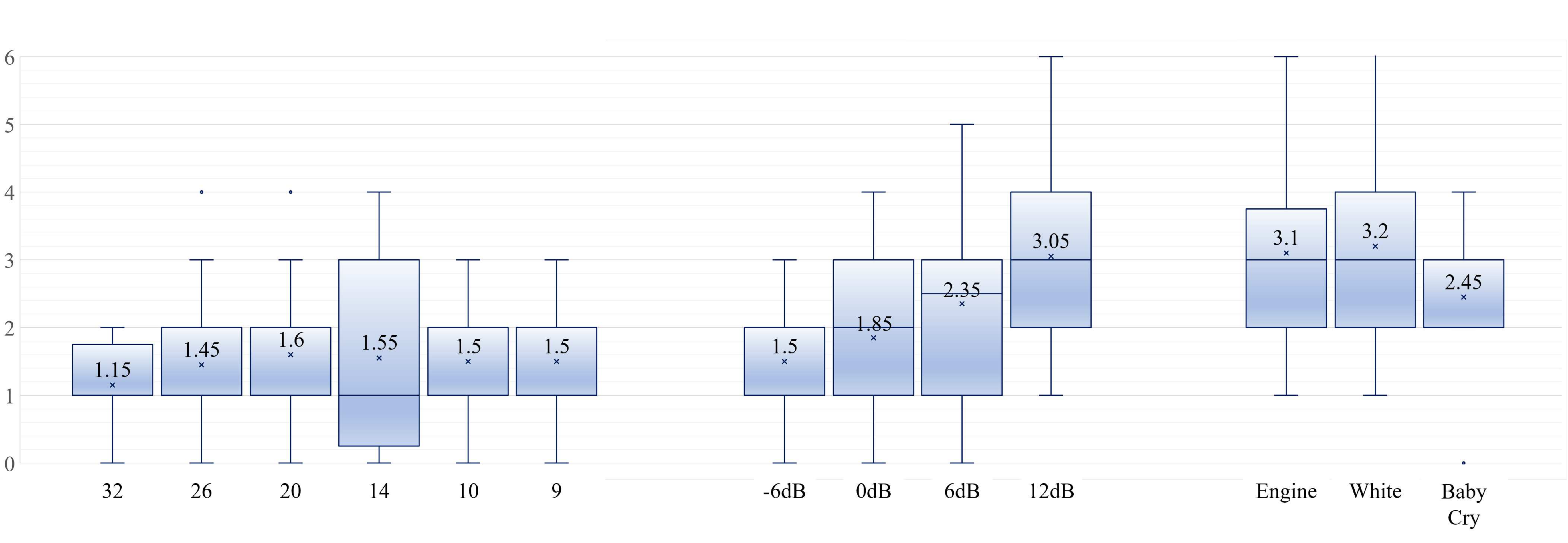}
        \label{fig:JND_c}
    }
	\caption{
	Statistic box plots of the response times and the amounts of DIFF responses of the 20 participants in the 100-pair enhanced utterances A/B Test.
	(a) the amount of DIFF responses;
	(b) the average response times of total 100 pairs and the pairs of DIFF responses are in units of milliseconds;
	(c) the amount of DIFF responses under the SEOFP-NETs with 6 different bit-widths, 4 different SNR levels, and 3 different noise types.
	}
	\label{fig:JND_exp}
\end{figure*}

Figure~\ref{fig:JND_exp} illustrates the statistical boxplots of the average response times and the amount of time for the 20 participants to reach a DIFF response to the A/B test on the 100 pairs of enhanced utterances; these response times are in milliseconds, as shown in Figure~\ref{fig:JND_b}. From Figure~\ref{fig:JND_a} and~\ref{fig:JND_b}, the average time of the 100 responses is 746.36 ms. This short response time indicates that the JND statistic may be applied. In addition, the average quantity of DIFF responses is only 8.75 out of the 100 pairs of enhanced utterances, and the average time of DIFF responses is 915.09 ms. These results have two important implications. First, although reductions in the scores of the standardized objective evaluation metrics (i.e., PESQ and STOI) are observed, as summarized in Tables~\ref{tab:de-noise}, in most pairs of utterances, the two enhanced speech signals sound similar to the listeners. Second, the longer average time of DIFF responses than the average response time of the 100 pairs indicated that the listeners were unable to effortlessly differentiate between the enhanced speech signals from the baseline model and the proposed SEOFP-NETs although they ultimately made the DIFF decision.


To further probe into the scenarios of DIFF responses, the number of DIFF responses under different conditions are categorized into three groups, as shown in Figure~~\ref{fig:JND_c}. The first group consisted of SEOFP-NETs with six different bit widths; the second group had four different SNR levels (-6, 0, 6, and 12); and the last group had three different noise types (engine, white, and baby cry). From the figure, the 8.75 DIFF responses are observed to be evenly distributed among the six different bit widths of the SEOFP-NETs even though the interquartile ranges are slightly different among these models. However, note that the 32-bit model has an average of 1.15 DIFF responses although the A/B enhanced speech signals are actually the same. These erroneous evaluations indicate that the participants were extremely unsure in making the DIFF decisions.


For the different SNR levels, 12 dB is observed to receive the most quantity of DIFF responses, whereas -6 dB had the least quantity among the 100 pairs of enhanced speech signals. A possible reason for this result is that the distortions of the baseline model and SEOFP-NETs are different. The original noisy speech signals with high SNRs inherently have high quality and intelligibility. However, the distortion of enhanced noisy speech signals with high SNRs is more evident than the distortion of enhanced noisy speech signals with low SNRs. Moreover, the various models may cause different distortions while enhancing the noisy speech signals; consequently, the differentiation made by the participants between the two speech signals is based on different distortions. For the different noise types, the average quantity of DIFF responses to the $engine$ noise is 3.1; this approximates 3.2, which is the average quantity of DIFF responses to the $white$ noise. In contrast, the \emph{Baby Cry} noise received the least quantity of DIFF responses among the 100 pairs of enhanced speech signals. These results indicate that a pair of enhanced speech signals with stationary noise can be more facilely differentiated by listeners than a pair with non-stationary noise.


\section{Conclusions}\label{sec:conclusion}

In this paper, a novel SEOFP-NET strategy is proposed to compress the model size and accelerate the inference time for speech enhancement tasks of regression in speech signal processing.
In the offline training phase, the proposed SEOFP-NET compressed the sizes of the DNN models by quantizing the fraction bits of single-precision floating-point parameters. Before the online inference is implemented, all parameters in the trained SEOFP-NET model are slightly adjusted to accelerate the inference time by replacing the floating-point multiplier logic circuit with an integer-adder logic circuit. For generalization, the proposed SEOFP-NET technique is applied to the two important speech enhancement tasks of regression in speech signal processing, i.e., speech denoising and speech dereverberation, with two different model architectures (BLSTM and FCN) under two common corpora (TIMIT and TMHINT).
The experimental results show that the SEOFP-NET models compared with the baseline models can be significantly compressed between 71.859\% and 81.249\% in terms of size and accelerated between 1.192$\times$ and 1.212$\times$ with respect to the inference time, respectively.
Moreover, the speech denoising and dereverberation performance achieved by the SEOFP-NET models is similar to that of the baseline models in PESQ and STOI, which are the standardized objective evaluation metrics.
The results also indicate that the proposed SEOFP-NET can cooperate with other efficiency strategies to achieve a synergy effect.
In addition, the JND in the user study experiment is employed to statistically analyze the effect on listening. The results indicate that the listeners cannot facilely differentiate between the enhanced speech signals from the baseline model and the proposed SEOFP-NETs. To the best knowledge of the authors, this study is one of the first research works to substantially reduce the size of DNN-based algorithms and inference time, simultaneously, for speech enhancement tasks of regression in speech signal processing while maintaining satisfactory performance. The promising results suggest that the DNN-based speech enhancement algorithms with the proposed SEOFP-NET technique can be suitably applied to lightweight embedded devices.

\ifCLASSOPTIONcaptionsoff
  \newpage
\fi



%

%


\bibliographystyle{ieeetr}
\bibliography{refs}




\end{document}